\documentclass[sigconf,screen]{acmart}
\renewcommand\footnotetextcopyrightpermission[1]{} 
\AtBeginDocument{%
  }

\setcopyright{acmlicensed}
\copyrightyear{2018}
\acmYear{2018}
\acmDOI{XXXXXXX.XXXXXXX}
\acmConference[Conference acronym 'XX]{Make sure to enter the correct
  conference title from your rights confirmation email}{June 03--05,
  2018}{Woodstock, NY}
\acmISBN{978-1-4503-XXXX-X/2018/06}

\usepackage{enumitem}
\usepackage{bbm}
\usepackage{multirow}
\usepackage{subcaption}
\usepackage{graphicx}
\DeclareMathOperator*{\argmax}{arg\,max}

\begin{document}

\title{Personalized Pricing in Social Networks with Individual and Group Fairness Considerations}

\author{Zeyu Chen}
\affiliation{%
  \institution{University of Delaware}
  \city{Newark}
  \state{DE}
  \country{USA}
}
\email{chenze@udel.edu}

\author{Bintong Chen}
\authornote{Corresponding Authors.}
\affiliation{%
  \institution{University of Delaware}
  \city{Newark}
  \state{DE}
  \country{USA}}
\email{bchen@udel.edu}

\author{Wei Qian}
\authornotemark[1]
\affiliation{%
  \institution{University of Delaware}
  \city{Newark}
  \state{DE}
  \country{USA}}
\email{weiqian@udel.edu}

\author{Jing Huang}
\affiliation{%
  \institution{Emory University}
  \city{Atlanta}
  \state{GA}
  \country{USA}}
\email{jing.huang@emory.edu}


\begin{abstract}
Personalized pricing assigns different prices to customers for the same product based on customer-specific features to improve retailer revenue.
However, this practice often raises concerns about fairness at both the individual and group levels. At the individual level, a customer may perceive unfair treatment if he/she notices being charged a higher price than others. At the group level, pricing disparities can result in discrimination against certain protected groups, such as those defined by gender or race.
Existing studies on fair pricing typically address individual and group fairness separately. This paper bridges the gap by introducing a new formulation of the personalized pricing problem that incorporates both dimensions of fairness in social network settings.
To solve the problem, we propose FairPricing, a novel framework based on graph neural networks (GNNs) that learns a personalized pricing policy using customer features and network topology.
In FairPricing, individual perceived unfairness is captured through a penalty on customer demand, and thus the profit objective, while group-level discrimination is mitigated using adversarial debiasing and a price regularization term.
Unlike existing optimization-based personalized pricing, which requires re-optimization whenever the network updates, the pricing policy learned by FairPricing assigns personalized prices to all customers in an updated network based on their features and the new network structure, thereby generalizing to network changes.
Extensive experimental results show that FairPricing achieves high profitability while improving individual fairness perceptions and satisfying group fairness requirements.
\end{abstract}





\maketitle

\section{Introduction}
In recent years, firms across various industries have increasingly adopted personalized pricing to improve profitability, with notable examples including grocery retailers \cite{Clifford2012shopper}, online employment marketplace \cite{Wallheimer2018are}, and travel websites \cite{Mattioli2012on}. Instead of offering uniform prices, firms leverage customer features to estimate each individual’s willingness to pay and determine a customized price for them accordingly. In practice, such customization often takes the form of personalized discount offers relative to a publicly announced baseline price \cite{elmachtoub2021value}. Meanwhile, the rapid expansion of web-based social networking platforms like TikTok and Facebook has established an important new channel for customer engagement and product promotion \cite{anderson2011turning, rachmad2022social}. These platforms provide rich individual-level demographic, behavioral, and relational data, making them particularly suitable for implementing personalized pricing.

However, due to its potential discriminatory effects, personalized pricing has raised important concerns about fairness \cite{van2020note, miller2014we, hutchinson2022challenges}, which can be broadly categorized into two types. At the individual level, customers may perceive unfairness if they realize they are being charged more than others for the same product \cite{xia2010good, jin2014power}, particularly within social networks where pricing information can be easily shared among neighbors. Such perceptions can reduce trust in the retailer, weaken customer loyalty, and lower purchase intentions \cite{garbarino2003dynamic, rotemberg2011fair, richards2016personalized}. At the group level, the focus shifts to demographic groups defined by protected attributes such as race, gender, and ethnicity. For example, women face higher prices for personal care products \cite{moshary2023gender}, and some racial groups are charged more for test preparation services \cite{Angwin2015the}. These price gaps are harmful and even illegal because they create systematic disparities against protected groups \cite{fang2020gender, butler2023racial}. Unlike individual-level fairness, which is directly linked to customer perceptions, group-level fairness is typically enforced through government regulations. Examples include New York State Senate Bill S2679 that bans gender-based differential pricing, and price discrimination monitoring undertaken by the Federal Trade Commission \cite{FTC2025price}. To address these fairness concerns, recent research has emphasized the development of fair personalized pricing methods \cite{xu2022regulatory, das2022individual, chen2023personalized, chen2025fairness, zhang2022peer, kallus2021fairness}. Depending on the level of fairness considered, these methods can be classified into individual fairness methods and group fairness methods.

For individual fairness, most research focuses on personalized pricing in social networks \cite{bloch2013pricing, nie2026opaque, alon2013differential, amanatidis2021inequity}, where connected customers can compare prices with one another. For example, one stream of research \cite{alon2013differential, amanatidis2021inequity, chopra2022extended} requires connected customers to receive similar prices, while another \cite{bloch2013pricing, duan2021optimal, nie2026opaque} incorporates the price differences a customer realizes when comparing with neighbors as a penalty term on customer demand.
These methods determine personalized prices by solving an optimization problem under a given network structure rather than learning a feature-based pricing policy. They require re-optimization when the network changes (e.g., when new users or connections are added), which may limit their practical applicability.
As for group fairness, existing studies impose fairness constraints to ensure that a personalized pricing policy provides more equitable treatment across demographic groups. These constraints typically involve enforcing similar levels of demand and consumer surplus across groups \cite{cohen2022price}, or bounding the distance between their price distributions \cite{chen2023personalized, kallus2021fairness}.
While both individual and group fairness are crucial in personalized pricing, prior research has largely addressed them separately.

Therefore, a natural question arises: \textit{can we design a personalized pricing framework that increases retailer profitability while improving individual fairness perceptions and complying with regulatory group fairness requirements?} To answer the question, in this paper, we first introduce a novel problem formulation of the personalized pricing in social networks with both individual and group fairness considerations. In social network settings, the price comparison among connected neighbors generates perceptions of unfairness. Our formulation captures the impact of these perceptions on customer demand, and models the retailer's task as a profit maximization problem subject to group fairness constraints.
To solve this problem, we propose a novel graph neural network (GNN)-based framework named \textbf{FairPricing} for fair personalized pricing. In the framework, the GNN-based pricing policy is trained to determine the optimal price for each node based on node representations learned from both node features and topological structures. To mitigate the impact of protected attributes on  pricing decisions, we incorporate an adversarial learning module that debiases the representations, a technique widely adopted in fair machine learning \cite{zhang2018mitigating, wadsworth2018achieving, xu2019achieving}. We also add a regularization term on the output prices to further stabilize group fairness performance. Notably, unlike optimization-based methods that require re-optimization whenever the network changes, our learned feature-based pricing policy generalizes to such changes. We summarize our contributions as follows:
\begin{itemize}[topsep=5pt]
    \item To the best of our knowledge, this is the first work to formulate the personalized pricing problem in social networks that incorporates both individual and group fairness.
    \item We propose FairPricing, a novel GNN-based framework that trains personalized pricing policies to optimize retailer profit while improving individual fairness perceptions and satisfying group fairness requirements, with the trained pricing policy readily generalizable to network changes.
    \item Experimental results on different datasets demonstrate the effectiveness of our method in addressing fairness concerns in personalized pricing and increasing retailer profitability. Meanwhile, the trained policy generalizes well under moderate network changes. We further conduct post-hoc analyses of the policy and offer useful managerial insights. 
\end{itemize}

\section{Related Work}
In this section, we review two research streams that are most relevant to this paper: fair personalized pricing methods and graph neural networks.

\subsection{Fair Personalized Pricing Methods}

\paragraph{Individual fairness methods}
This category of methods addresses the unfairness perceived by customers when they realize that they are charged different prices for the same product \cite{martins1995experimental, jin2014power, xia2010good}.
\citet{das2022individual} study a setting where customers are partitioned into segments, and the retailer sets segment-level prices. Their approach aims to bound the price difference between any two segments by their feature dissimilarity.
Another work \cite{xu2022regulatory} focuses on the maximum and minimum prices set by the retailer, introducing constraints that restrict both their difference and their ratio.
More closely related to our study, a line of research centers on personalized pricing in social networks, where perceived unfairness arises from price differences among connected customers \cite{bloch2013pricing, nie2026opaque, alon2013differential, amanatidis2021inequity}.
For each edge in a network, the fairness constraints in \cite{alon2013differential} require the price difference between the two connected nodes to be bounded. They show that such constraints are too restrictive, leading to nearly uniform pricing in most networks, and then propose a relaxation that allows to choose certain nodes to exclude from pricing (i.e., their edges then carry no constraints). Under this relaxation, the profit maximization problem with fairness constraints is proven to be NP-hard in general \cite{amanatidis2021inequity}, which motivates the development of several approximation algorithms \cite{amanatidis2021inequity, chopra2022extended}.
In contrast to explicitly imposing a constraint on each edge, an alternative method \cite{bloch2013pricing} models the difference between the price for a node and the average price of its neighbors as a penalty term in the node utility function. This penalty affects customer utility and consequently their demand. Customers determine their demand by maximizing utility given the prices, while the retailer chooses prices to maximize profit based on customer demand. The equilibrium of this game yields the optimal personalized prices. Subsequent studies extend this formulation to incorporate consumption network effects \cite{duan2021optimal} and to investigate the impact of opaque personalized pricing strategies \cite{nie2026opaque}.

\paragraph{Group fairness methods}
This category of methods mitigates potential discrimination against protected groups in personalized pricing, aiming to ensure that different demographic groups receive equitable treatment in compliance with consumer protection regulations \cite{chen2023personalized, cohen2022price, kallus2021fairness, cohen2025dynamic, yang2024fairness, xu2023doubly, chen2025fairness}. The family of fairness measures in \cite{cohen2022price} requires group differences to be bounded in terms of price, demand, consumer surplus, and social welfare. Their method is based on a setting where all customers within a group are assigned the same price. Moving beyond the simplified setting, later approaches determine personalized prices based on customer features while imposing constraints that bound the distance between the price distributions of two groups \cite{chen2023personalized, kallus2021fairness}. Some studies further extend to dynamic pricing scenarios, introducing constraints that restrict the price differences between groups in each time period \cite{cohen2025dynamic, chen2025fairness, xu2023doubly}. While these methods generally assume a monopolist selling a single product, several recent works also investigate group fairness in settings with duopoly competition \cite{yang2024fairness} or multi-product sales \cite{wang2025data}.

Despite extensive research in this field, existing methods typically address individual and group fairness separately. Therefore, we study the novel problem of personalized pricing in social networks that incorporates both dimensions of fairness. 

\subsection{Graph Neural Networks}
Graph neural networks (GNNs) have emerged as a powerful tool for learning from graph data, and have been successfully applied to tasks such as recommender systems \cite{gao2023survey}, fraud detection \cite{cheng2025graph}, and drug discovery \cite{bongini2021molecular}. GNNs are commonly categorized into two main classes \cite{wu2020comprehensive}: spectral-based and spatial-based methods. Spectral-based GNNs, first introduced by \citet{bruna2013spectral}, define graph convolution using the graph Fourier transform derived from the eigen-decomposition of the graph Laplacian. Subsequent studies propose various extensions, leading to a broader family of spectral-based methods \cite{henaff2015deep, defferrard2016convolutional, kipf2017semisupervised, levie2018cayleynets, zhuang2018dual}. Graph Convolutional Network (GCN) \cite{kipf2017semisupervised}, one of the most popular extensions, simplifies the graph convolution operation by employing a first-order polynomial approximation, thereby reducing computational complexity. Alternatively, spatial-based GNNs \cite{hamilton2017inductive, velivckovic2018graph, zhang2018end, chen2018fastgcn, chen2018stochastic} define graph convolution in the spatial domain by directly updating the node representation using the aggregated information from its local neighborhood. Representative examples include Graph Attention Network (GAT) and GraphSAGE. GAT \cite{velivckovic2018graph} utilizes self-attention in neighborhood aggregation by assigning learnable attention weights to different neighbors, enabling more expressive message passing. GraphSAGE \cite{hamilton2017inductive} samples a fixed-size set of neighbors and aggregates information from the sampled neighborhood, which allows mini-batch training that improves scalability and supports inductive learning. Recent studies highlight that GNNs can suffer from fairness issues, motivating efforts to develop fair GNNs to mitigate biases \cite{dai2021say, guo2023towards, zhang2023fpgnn, zhu2024fairagg, dong2021individual}. FairGNN \cite{dai2021say} integrates adversarial debiasing with a fairness regularizer that penalizes correlations between predictions and protected attributes, thereby addressing group fairness. Complementarily, REDRESS \cite{dong2021individual} defines individual fairness from a ranking perspective and aims to preserve the similarity order of instances between the input and outcome spaces.

Existing fair GNN studies focus on classical graph learning tasks like node classification and link prediction, while our work is different as it proposes a GNN-based framework specifically designed to address individual and group fairness in personalized pricing.

\section{Problem Formulation}
We first introduce the personalized pricing setting in social networks and define the individual and group fairness considerations within this context. Then, we formulate the personalized pricing problem that incorporates both dimensions of fairness.

\subsection{Personalized Pricing in Social Networks}
We consider a market where a monopolist retailer sells a single product with unlimited supply to customers.
Let $\mathcal{G}=(\mathcal{V},\mathcal{E},\mathbf{X},\mathbf{S})$ denote the social network formed by customers, where $\mathcal{V}=\{v_1,\dots,v_n\}$ is the set of $n$ customers, $\mathcal{E} \subseteq \mathcal{V} \times \mathcal{V}$ represents the set of edges connecting them, $\mathbf{X}=\{\mathbf{x}_1,\dots,\mathbf{x}_n\}$ denotes the non-protected attributes of customers, and $\mathbf{S}=\{s_1,\dots,s_n\}$ denotes the protected attributes.
Define the adjacency matrix of the network $\mathcal{G}$ as $\mathbf{A} \in \mathbb{R}^{n \times n}$, where $a_{ij}=1$ if customers $v_i$ and $v_j$ are connected and $a_{ij}=0$ otherwise, with $a_{ii}=0$ for all $i$.
The $m$-dimensional non-protected attributes $\mathbf{x}_i \in \mathbb{R}^m$ and the protected attribute $s_i$ together characterize each customer. For technical brevity, we assume $s_i \in \{0,1\}$ is binary, although our framework can be straightforwardly extended to multi-class cases. The customer feature vector $\tilde{\mathbf{x}}_i \coloneq (\mathbf{x}_i, s_i) \in \mathbb{R}^m \times \{0,1\}$ is observable to the retailer.

Each customer has a willingness to pay $u_i$ for the product, which follows a distribution with cumulative distribution function $F_{\tilde{\mathbf{x}}_i}(\cdot)$ that depends on the customer features and is assumed to be known by the retailer.
Under a personalized pricing policy, if the $i$th customer is assigned a price $p_i$, he or she purchases the product only if $u_i \geq p_i$. Thus, the demand is $\mathbbm{1}(u_i \geq p_i)$, where $\mathbbm{1}(\cdot)$ denotes the indicator function. The $i$th customer's expected demand given $p_i$ is
\begin{equation}
\label{eqn:expdemand}
    d_i(p_i) = \mathrm{E}_{u_i \sim F_{\tilde{\mathbf{x}}_i}(\cdot)} [\mathbbm{1}(u_i \geq p_i)] = 1 - F_{\tilde{\mathbf{x}}_i}(p_i).
\end{equation}
For example, under the popular linear demand model \cite{chen2023personalized, cohen2025dynamic, cohen2022price, xu2022regulatory}, $u_i \sim U(0, g(\tilde{\mathbf{x}}_i))$ and $d_i(p_i) = \max\{0,1-p_i/g(\tilde{\mathbf{x}}_i)\}$, where $g(\cdot)$ is a function known by the retailer.

Let $c > 0$ denote the marginal cost of the product. The expected profit from the $i$th customer is $\pi_i(p_i) \coloneq (p_i - c) d_i(p_i)$. Without any fairness consideration, the total expected profit is maximized by setting the personalized prices as $p_i^* = \argmax_{p_i}\pi_i(p_i)$ for $i=1,\dots,n$. 
However, in practice, individual perceptions of unfairness reduce customer purchase intentions, and price discrimination against protected groups is not allowed by regulations. Therefore, a personalized pricing policy in social networks should incorporate both fairness considerations, which we define formally in Section~\ref{subsec:fairmeasure}.

\subsection{Individual Fairness and Group Fairness} \label{subsec:fairmeasure}
We begin this section by defining individual fairness measures. Under personalized pricing, customers perceive unfairness if they are charged a price higher than a reference price, which places them at a disadvantage \cite{jin2014power, xia2004price}. In social networks, connected customers can readily share pricing information, and a common choice of reference price is the average price of a customer’s neighbors \cite{duan2021optimal, bloch2013pricing, colombo:hal-05090073, liu2025optimal}. With this reference, the realized price difference is given by
\begin{equation}
\label{eqn:pricedifference}
    \Delta_i = 
    \begin{cases}
    0 & r_i=0 \\
    p_i - \frac{1}{r_i} \sum_{j=1}^n a_{ij} p_j & r_i \neq 0
    \end{cases},
\end{equation}
where $r_i = \sum_{j=1}^n a_{ij}$ is the total number of neighbors for the $i$th customer. When $r_i = 0$, customer $i$ is isolated and no price comparison is performed.
$\Delta_i$ induces perceptions of price unfairness. A positive $\Delta_i$ indicates that the customer pays more than the reference price, triggering perceived unfairness; conversely, the customer pays less with a negative $\Delta_i$, placing them in an advantageous position that leads to positive emotions. Based on this intuition, we propose the following measure of perceptions of unfairness.

\begin{definition}[Perceptions of Price Unfairness]
    Given a network $\mathcal{G}=(\mathcal{V},\mathcal{E},\mathbf{X},\mathbf{S})$ and a personalized pricing policy $\mathbf{p}=(p_1,\dots,p_n)$ on $\mathcal{G}$ that assigns a price $p_i$ to each customer $v_i \in \mathcal{V}$, the perception of price unfairness for $v_i$ under policy $\mathbf{p}$ is defined as
    \begin{equation} \label{eqn:unfairnessperception}
        \eta_i(\mathbf{p}) =
        \begin{cases}
            \tanh(\alpha \Delta_i) & \Delta_i \leq 0 \\
            \tanh(\beta \Delta_i) & \Delta_i > 0
        \end{cases},
    \end{equation}
    where $\Delta_i$ is defined in \eqref{eqn:pricedifference}, $\tanh(x) = (e^x - e^{-x}) / (e^x + e^{-x})$, and $0 < \alpha < \beta$.
\end{definition}

Here, $\eta_i(\mathbf{p})$ measures individual perceived unfairness by mapping the unbounded $\Delta_i$ into $[-1,1]$. We use the hyperbolic tangent function $\tanh(\cdot)$ for this mapping as it preserves the sign of $\Delta_i$ and increases monotonically with $\Delta_i$. These desirable properties also make $\tanh(\cdot)$ widely used in other domains, such as defining biometric matching scores \cite{jain2005score} and modeling corporate risk-aversion factors \cite{lerche1999economic}.
$\alpha$ and $\beta$ capture customer sensitivity to price differences, and the asymmetry $\alpha < \beta$ reflects that customers are more sensitive when disadvantaged than when advantaged by the difference, as explained by prospect theory \cite{kahneman2013prospect}. These parameters are assumed to be known by the retailer.
Then, we consider the impact of unfairness perceptions on customer demand and define the unfairness-adjusted expected demand of customer $i$ under policy $\mathbf{p}$:
\begin{equation}
\label{eqn:adjexpdemand}
    \bar{d}_i(\mathbf{p}) = (1-\eta_i(\mathbf{p})) d_i(p_i),
\end{equation}
where $d_i(p_i)$ is the baseline demand in \eqref{eqn:expdemand}. For $\bar{d}_i(\mathbf{p})$, perceived unfairness reduces customer demand, while positive emotions from receiving favorable deals increase it, which is consistent with customer behavior discussed in prior work \cite{duan2021optimal, bloch2013pricing, lu2016joint, hu2016dynamic, liu2025optimal}.

Next, we introduce group fairness measures. Group fairness requires demographic groups defined by protected attributes to be treated equitably. In the context of personalized pricing with binary protected attributes $s_i \in \{0,1\}$, similar to \cite{kallus2021fairness}, we consider a group fairness criterion measuring the difference between the average prices of the two groups, which is defined as follows.

\begin{definition}[Group-level Price Fairness] \label{def:group-levelfair}
    For customers with protected attributes $\mathbf{S}=\{s_1,\dots,s_n\}$, a personalized pricing policy $\mathbf{p}=(p_1,\dots,p_n)$ satisfies group-level price fairness up to threshold $\tau \geq 0$ if
    \begin{equation}
    \label{eqn:groupfairness}
        \left| \frac{1}{n_0} \sum_{i: s_i=0} p_i - \frac{1}{n_1} \sum_{i: s_i=1} p_i \right| \leq \tau,
    \end{equation}
    where $n_0$ and $n_1$ denote the sizes of the two groups, respectively.
\end{definition}

The threshold $\tau$ is the maximum tolerance for the difference in average prices between the groups (e.g., set by regulatory requirements). This measure ensures that group-level disparities remain within an acceptable bound, aligning with anti-discrimination laws.

\subsection{Problem Definition}
For a network $\mathcal{G}$, we define a pricing policy $\rho_{\mathbf{w}}: (\mathbf{X}, \mathbf{A}) \mapsto \mathbf{p} \in \mathbb{R}_+^n$ that maps customer features and topological structures to personalized prices, where $p_i = (\rho_{\mathbf{w}}(\mathbf{X}, \mathbf{A}))_i$ and $\mathbf{w}$ denotes the learnable parameters. Protected attributes $\mathbf{S}$ are excluded from the input, as their direct use is typically not allowed by regulations \cite{veale2017fairer, hellman2020measuring}.
In contrast to optimization-based methods that directly compute optimal $p_i$ values for each customer in a given network, which results in solutions valid only for that network, a learned feature-based pricing policy $\rho_{\mathbf{w}}(\mathbf{X}, \mathbf{A})$ naturally generalizes to network changes by assigning prices to all customers in an updated network using their features and the new network structure as input.

With the fairness measures in Section~\ref{subsec:fairmeasure}, when determining an optimal personalized pricing policy $\rho_{\mathbf{w}}(\mathbf{X}, \mathbf{A})$ to maximize profit, a retailer should comply with group fairness requirements imposed by regulations while mitigating customer perceptions of unfairness.
We now formally define the problem of personalized pricing in social networks with individual and group fairness considerations.

\begin{definition}[Fair Personalized Pricing in Social Networks]
    Given a network $\mathcal{G}=(\mathcal{V},\mathcal{E},\mathbf{X},\mathbf{S})$ and a threshold $\tau$ on the group fairness measure in \eqref{eqn:groupfairness}, determine a personalized pricing policy $\rho_{\mathbf{w}}(\mathbf{X}, \mathbf{A})$ for $\mathcal{G}$ under fairness considerations by solving the problem:
    \begin{equation} \label{eqn:fairpricingdef}
    \begin{aligned}
        \max_{\rho_{\mathbf{w}}(\cdot, \cdot)} \;\;& \sum_{i=1}^n \left[(\rho_{\mathbf{w}}(\mathbf{X}, \mathbf{A}))_i - c \right] \bar{d}_i(\rho_{\mathbf{w}}(\mathbf{X}, \mathbf{A})) \\
        \text{s.t.} \;\;& \left| \frac{1}{n_0} \sum_{i: s_i=0} (\rho_{\mathbf{w}}(\mathbf{X}, \mathbf{A}))_i - \frac{1}{n_1} \sum_{i: s_i=1} (\rho_{\mathbf{w}}(\mathbf{X}, \mathbf{A}))_i \right| \leq \tau
    \end{aligned},
    \end{equation}
    where $\left[(\rho_{\mathbf{w}}(\mathbf{X}, \mathbf{A}))_i - c \right] \bar{d}_i(\rho_{\mathbf{w}}(\mathbf{X}, \mathbf{A}))$ denotes the expected profit contribution from customer $i$, with demand adjusted for perceived unfairness as defined in \eqref{eqn:adjexpdemand}.
\end{definition}

\section{Methodology}
\begin{figure}[htbp]
  \centering
  \includegraphics[width=0.9\columnwidth]{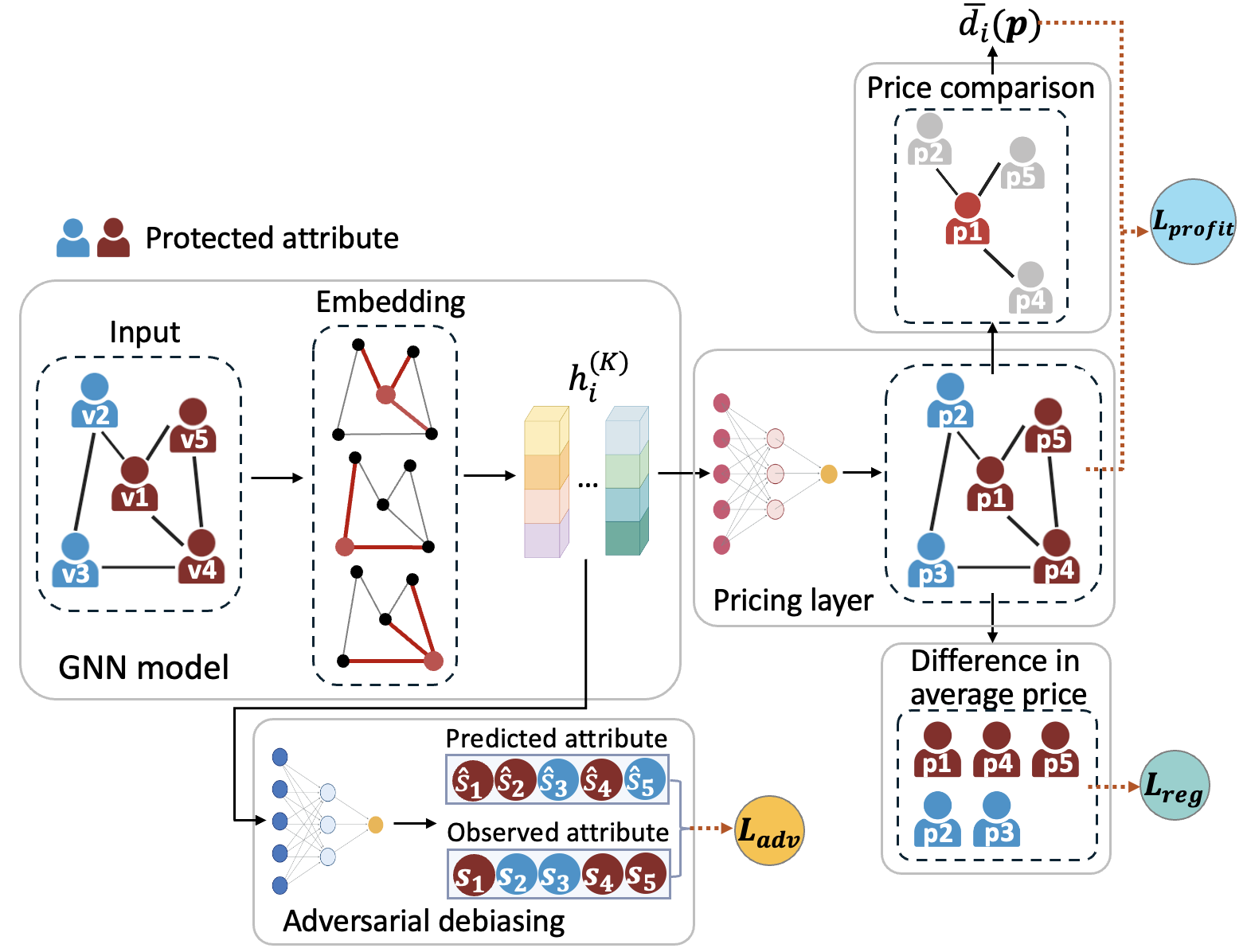}
  \caption{The framework of FairPricing.}
  \Description{This is a flowchart.}
  \label{fig:flowchart}
\end{figure}

In this section, we propose FairPricing, a GNN-based framework designed to solve the problem in \eqref{eqn:fairpricingdef} by determining the optimal pricing policy $\rho_{\mathbf{w}}(\mathbf{X}, \mathbf{A})$. 
The overall architecture of FairPricing is illustrated in Figure~\ref{fig:flowchart}, which consists of a GNN pricing module $\rho$ and an adversarial debiasing module $\psi_A$.
In the pricing module $\rho$, a GNN model $\psi_G$ takes $(\mathbf{X}, \mathbf{A})$ as input to learn node representations, which are subsequently fed into a linear pricing layer $\psi_P$ that maps each representation vector to a personalized price, hence $\rho = \psi_P \circ \psi_G$. The resulting prices determine customer demand adjusted for perceived unfairness, which in turn drives retailer profit.
The adversarial debiasing module introduces an adversary $\psi_A$ that seeks to predict the protected attribute from the representations produced by $\psi_G$, while $\psi_G$ is trained to fool the adversary by learning fair representations that eliminate protected attribute information.
We provide theoretical results showing that this minimax game yields fair representations with respect to the protected attribute, leading to personalized prices that do not discriminate against protected groups.
In addition, to prevent potential violations of group fairness regulations caused by the instability in adversarial training, we directly incorporate the group fairness constraint in \eqref{eqn:fairpricingdef} as a regularization term on the prices given by $\psi_P$ that improves the regulatory compliance of the pricing policy.
In the remainder of this section, we introduce each component in detail.

\subsection{The Pricing Module}
For a given network $\mathcal{G}$, the pricing module employs a GNN model $\psi_G: (\mathbf{X}, \mathbf{A}) \mapsto \mathbf{H} \in \mathbb{R}^{n \times q}$ to learn a $q$-dimensional representation vector $\mathbf{h}_i$ for each $v_i \in \mathcal{V}$ based on its attributes and the network topology.
GNNs learn these representations through iterative neighborhood aggregation, where at each layer, the representation of a node is updated by combining its previous representation with the information aggregated from its neighboring nodes. After $k$ layers, each node representation encodes the structural information of its $k$-hop neighborhood. This update process can be formalized as
\begin{equation}
    \begin{aligned}
    \mathbf{z}_i^{(k)} &= \mathrm{AGG}^{(k)} \left( \left\{\mathbf{h}_j^{(k-1)}: \, v_j \in \mathcal{N}(v_i) \right\} \right) \\
    \mathbf{h}_i^{(k)} &= \mathrm{COMB}^{(k)} \left( \mathbf{h}_i^{(k-1)}, \mathbf{z}_i^{(k)} \right),
    \end{aligned}
\end{equation}
where $\mathbf{h}_i^{(k)}$ denotes the representation vector of node $v_i$ at the $k$th layer, $\mathcal{N}(v_i)$ represents the set of its neighboring nodes, and $\mathrm{AGG}^{(k)}(\cdot)$ and $\mathrm{COMB}^{(k)}(\cdot)$ are the aggregation and combination functions at layer $k$, respectively, whose specific forms vary across different GNN architectures. In the proposed FairPricing framework, any GNN architecture following the above update process can be adopted, including GCN \cite{kipf2017semisupervised}, GAT \cite{velivckovic2018graph}, and GraphSAGE \cite{hamilton2017inductive}.

With a $K$-layer $\psi_G$, the final representation $\mathbf{h}_i^{(K)}$ of $v_i$ is obtained from the model output $\psi_G(\mathbf{X}, \mathbf{A}; \mathbf{w}_G)$, where $\mathbf{w}_G$ are the learnable parameters. The representations are then passed to a linear pricing layer $\psi_P: \mathbb{R}^q \mapsto \mathbb{R}_+$, which maps $\mathbf{h}_i^{(K)}$ to a personalized price $p_i$ as
\begin{equation}
    p_i = p_{\textrm{max}} \sigma\left(\mathbf{w}_P^T \mathbf{h}_i^{(K)} + b_P \right),
\end{equation}
where $\mathbf{w}_P$ and $b_P$ are learnable parameters, $\sigma(\cdot)$ denotes the sigmoid function, and $p_{\textrm{max}}$ specifies the upper pricing bound determined by the retailer. This ensures that each price $p_i$ lies within the admissible range $[0, p_{\textrm{max}}]$. The personalized pricing policy can therefore be written as $\rho_{\mathbf{w}}(\mathbf{X}, \mathbf{A}) = \psi_P \circ \psi_G(\mathbf{X}, \mathbf{A})$, where $\mathbf{w}=\{\mathbf{w}_G, \mathbf{w}_P, b_P\}$. To maximize profitability, the loss function for training $\rho_{\mathbf{w}}$ is defined as the negative average expected profit across all customers:
\begin{equation}
    \mathcal{L}_{\textrm{profit}} = -\frac{1}{n} \sum_{i=1}^n (p_i - c) \bar{d}_i(\mathbf{p}),
\end{equation}
where the profit calculation follows the objective function of \eqref{eqn:fairpricingdef}.

\subsection{The Adversarial Debiasing Module}
The prices assigned by $\psi_P$ may exhibit discrimination against the protected attribute because the representations from $\psi_G$ can encode biased information. These biases typically arise from correlations between protected and non-protected attributes and from group-related patterns within network structures, both of which can be amplified by the aggregation operation of GNNs. To mitigate such biases, recent studies have widely adopted adversarial learning as an effective approach for learning fair representations \cite{madras2018learning, feng2019learning, beutel2017data}.

In the adversarial debiasing module, an adversary $\psi_A$ receives the final-layer representation $\mathbf{h}_i^{(K)}$ from $\psi_G$ and predicts the protected attribute as $\hat{s}_i = \psi_A(\mathbf{h}_i^{(K)}; \boldsymbol{\theta}_A)$, where $\hat{s}_i$ is the predicted probability that $s_i = 1$ for $v_i$, and $\boldsymbol{\theta}_A$ denotes the parameters of $\psi_A$. The adversary $\psi_A$ aims to predict $s_i$ accurately, whereas $\psi_G$ learns to weaken the predictive power of $\psi_A$ by producing fair representations that eliminate biased information while preserving the signals essential for determining profitable prices in the downstream pricing layer $\psi_P$. This adversarial training process defines a minimax game between $\psi_G$ and $\psi_A$ with value function $V(\psi_G,\psi_A)$ as
\begin{equation} \label{eqn:advvaluationfunction}
\begin{aligned}
    \min_{\mathbf{w}_G} \max_{\boldsymbol{\theta}_A} \big\{ & V(\psi_G,\psi_A) \coloneq \mathrm{E}_{\mathbf{h}^{(K)} \sim f(\mathbf{h}^{(K)} | s = 1)} [\log(\psi_A(\mathbf{h}^{(K)}))] \\
    & + \mathrm{E}_{\mathbf{h}^{(K)} \sim f(\mathbf{h}^{(K)} | s = 0)} [\log(1-\psi_A(\mathbf{h}^{(K)}))] \big\},
\end{aligned}
\end{equation}
where $f(\mathbf{h}^{(K)} | s = 1)$ and $f(\mathbf{h}^{(K)} | s = 0)$ denote the conditional probability density functions of the learned representations $\mathbf{h}^{(K)}$ for customers in network $\mathcal{G}$ given the protected attribute $s=1$ and $s=0$, respectively. To show how the adversarial debiasing module improves fairness, we establish in the following proposition the relationship between the global optimum of this game and group-level price fairness with respect to the protected attribute.

\begin{proposition} \label{thm:advlearning}
The global minimum of the function $V(\psi_G,\psi_A)$ in \eqref{eqn:advvaluationfunction} is achieved if and only if $f(\mathbf{h}^{(K)} | s = 1) = f(\mathbf{h}^{(K)} | s = 0)$, $\forall \, \mathbf{h}^{(K)} \in \mathbb{R}^q$. At that point, the conditional distributions $f_{\psi_P}(\cdot|s)$ of the resulting prices $p=\psi_P(\mathbf{h}^{(K)})$ satisfy $f_{\psi_P}(p | s = 1) = f_{\psi_P}( p | s = 0)$, $\forall \, p \in [0, p_{\textrm{max}}]$, for any pricing layer function $\psi_P$.
\end{proposition}

The proof is provided in Appendix~\ref{appendixsec:proofofprop}. With this proposition, the global minimum of \eqref{eqn:advvaluationfunction} ensures that the prices given by $\psi_P$ have identical distributions across the two groups of $s$, thereby preventing group-level price discrimination. In practice, the objective function for implementing the adversarial debiasing module is
\begin{equation}
    \mathcal{L}_{\textrm{adv}} = -\frac{1}{n} \sum_{i=1}^n \left[s_i \log(\hat{s}_i) + (1-s_i) \log(1-\hat{s}_i) \right].
\end{equation}

\subsection{The FairPricing Framework}
The results in Proposition~\ref{thm:advlearning} indicate that, upon convergence of the adversarial debiasing process, the personalized price $p$ becomes independent of the protected attribute $s$. This independence condition represents a stronger form of group fairness, implying that the conditional distributions $p | s = 1$ and $p | s = 0$ have the equal mean value and satisfy the fairness criterion in \eqref{eqn:groupfairness}.
However, prior research has shown that adversarial training often suffers from instability issues, which can cause the training not to converge \cite{arjovsky2017towards, mescheder2018training, bang2018improved}. In our debiasing process, such non-convergence poses substantial risks for the retailer in terms of violating group fairness regulations. To mitigate these risks, we incorporate the constraint in \eqref{eqn:fairpricingdef} as a regularization term on the output prices of $\psi_P$, defined as
\begin{equation} \label{eqn:priceregularization}
    \mathcal{L}_{\textrm{reg}} =  \left| \frac{1}{n_0} \sum_{i: s_i=0} p_i - \frac{1}{n_1} \sum_{i: s_i=1} p_i \right|.
\end{equation}
This improves fairness compliance by explicitly encouraging equal average prices across the two groups, complementing the adversarial debiasing that enforces the stronger independence condition.

For the FairPricing framework, by integrating the pricing module, the adversarial debiasing module, and the additional regularization term above, the final objective function used for training is:
\begin{equation} \label{eqn:finalobjective}
    \min_{\mathbf{w}} \max_{\boldsymbol{\theta}_A} \; \mathcal{L}_{\textrm{profit}} + \lambda \mathcal{L}_{\textrm{reg}} - \phi \mathcal{L}_{\textrm{adv}}, 
\end{equation}
where $\mathbf{w}=\{\mathbf{w}_G, \mathbf{w}_P, b_P\}$ include the parameters of $\psi_G$ and $\psi_P$, and $\boldsymbol{\theta}_A$ corresponds to those of $\psi_A$. $\lambda$ and $\phi$ are two hyperparameters that control the strengths of the regularization term and the adversarial debiasing relative to the profit objective, respectively.
During the training, each iteration involves alternating updates between the parameters $\mathbf{w}$ and $\boldsymbol{\theta}_A$. Specifically, we first fix $\boldsymbol{\theta}_A$ and update $\mathbf{w}$ to minimize the overall objective function, and then fix $\mathbf{w}$ and update $\boldsymbol{\theta}_A$ to maximize the adversarial component in \eqref{eqn:finalobjective}. The optimization is performed using the Adam optimizer \cite{kingma2014adam}.

\section{Experiments}
In this section, we conduct numerical experiments utilizing two real-world datasets to evaluate the effectiveness of FairPricing for optimizing profit while addressing fairness concerns in personalized pricing. Specifically, we answer the following research questions.
\begin{itemize}[topsep=2pt]
    \item \textbf{RQ1}: Can FairPricing achieve high profitability while simultaneously improving individual fairness perceptions and complying with regulatory group fairness requirements?
    \item \textbf{RQ2}: Do the pricing policies learned by FairPricing generalize well to network changes, such as the addition of new users and connections?
    \item \textbf{RQ3}: How effective are the adversarial debiasing module and the price regularization term, and how do their hyperparameters in \eqref{eqn:finalobjective} affect the performance of FairPricing?
    \item \textbf{RQ4}: How do the personalized prices assigned by FairPricing relate to customer network connectivity, and what managerial insights does the relationship provide?
\end{itemize}

\subsection{Experimental Setup}
\subsubsection{Datasets}
We use two public datasets, \textbf{Pokec-z} and \textbf{Pokec-n} \cite{dai2021say, li2025fairness}, which are sampled from the anonymized 2012 Pokec social network data \cite{takac2012data} by selecting users from two different provinces.
Pokec is a popular online social networking platform in Slovakia.
Both datasets contain rich user profile information (e.g., gender, age, hobbies), and edges represent friendships between users. 
For each dataset, we consider a monopolist retailer selling a single product to the users in its network. We treat gender as the protected attribute and randomly select ten other user features as non-protected attributes; we assume that these attributes together determine customers’ willingness to pay for the product. The dataset statistics and attributes are summarized in Table~\ref{tab:datasetsummary} in Appendix~\ref{appendixsec:appendixdatasets}.

Following prior studies \cite{chen2023personalized, cohen2022price, chen2025fairness}, we consider two widely used demand models: the linear demand and the exponential demand. Under these models, the willingness to pay of a customer $u_i$ follows a uniform distribution $U(0, g(\tilde{\mathbf{x}}_i))$ and an exponential distribution $\textrm{Exp}(g(\tilde{\mathbf{x}}_i))$, respectively. Given a price $p_i$, the expected demand is $d_i(p_i) = \max\{0,1-p_i/g(\tilde{\mathbf{x}}_i)\}$ for the linear demand and $d_i(p_i) = \exp(-g(\tilde{\mathbf{x}}_i) p_i)$ for the exponential demand.
The function $g(\cdot)$ that determines the distribution parameter is known by the retailer. In our experiments, we conduct simulation studies by specifying the functional form of $g(\tilde{\mathbf{x}}_i)$. The detailed $g(\tilde{\mathbf{x}}_i)$ specifications for the two demand models in each dataset, along with the associated pricing bound $p_{\textrm{max}}$ and cost $c$ values, are provided in Appendix~\ref{appendixsec:appendixsimulation}.
In addition, the parameters $\alpha$ and $\beta$ in \eqref{eqn:unfairnessperception} are assumed to be known. For Pokec-z, we set $\alpha = 0.1$ and $\beta = 0.2$; for Pokec-n, we set $\alpha = 0.2$ and $\beta = 0.4$. This simulation setting models different levels of customer sensitivity to price differences across the two datasets.

\subsubsection{Metrics}
To evaluate the performance of personalized pricing policies, we consider three aspects: profitability, individual fairness perceptions, and group-level fairness. We measure profitability by the average profit
\begin{math}
    \pi_{\textrm{AVG}} = \frac{1}{n} \sum_{i=1}^n (p_i - c) \bar{d}_i.
\end{math}
Individual fairness is assessed by the average realized price difference $\Delta_{\textrm{AVG}} = \frac{1}{n} \sum_{i=1}^n \Delta_i$ and the average perception of price unfairness $\eta_{\textrm{AVG}} = \frac{1}{n} \sum_{i=1}^n \eta_i$. The definitions of $\bar{d}_i$, $\Delta_i$, and $\eta_i$ are introduced in Section~\ref{subsec:fairmeasure}. Finally, group-level discrimination is quantified according to Definition~\ref{def:group-levelfair} as
\begin{math}
    p_{\textrm{diff}} = |\frac{1}{n_0} \sum_{i: s_i=0} p_i - \frac{1}{n_1} \sum_{i: s_i=1} p_i|.
\end{math}

\subsubsection{Implementation Details} \label{subsubsec:implementation}
For FairPricing, to demonstrate its effectiveness under various GNN architectures, we implement three variants: \textbf{FairPricing-GCN}, \textbf{FairPricing-GAT}, and \textbf{FairPricing-GraphSAGE}, which use GCN \cite{kipf2017semisupervised}, GAT \cite{velivckovic2018graph}, and GraphSAGE \cite{hamilton2017inductive} as the GNN models for $\psi_G$, respectively.
To highlight the benefit of incorporating topological information through GNNs, we also implement \textbf{FairPricing-MLP} for comparison, which replaces $\psi_G$ with a multi-layer perceptron (MLP) that ignores network topology.
In addition, following previous works \cite{chen2023personalized, cohen2022price, bergemann2022third}, we include a \textbf{UniformPricing} policy that sets one uniform price across all customers as a benchmark for personalized pricing policies.

We randomly split each dataset into training/validation/testing sets with an 8/1/1 ratio.
In practice, a retailer typically trains a pricing policy on the existing network, and new users and connections may be added to the network later during policy deployment. To reflect this scenario, we train the model on the training set, using the combined training and validation sets for guiding early stopping in model training and the combined training and testing sets for final model evaluation. The validation and testing sets thereby represent newly added users and connections in the respective settings.

In our experiments, we set the group-level fairness threshold in \eqref{eqn:fairpricingdef} to $\tau = 0.5$. The UniformPricing method has $p_{\textrm{diff}} = 0$ and always satisfies this condition. For all other methods based on FairPricing, this threshold serves to identify valid candidate models that satisfy $p_{\textrm{diff}} \leq 0.5$ for model selection with early stopping. In addition, the hyperparameters $\lambda$ and $\phi$ of the FairPricing variants are determined by cross-validation with grid search for each demand model in each dataset. Furthermore, we run all experiments 5 times and report the average value and standard error of the evaluation metrics. We refer readers to Appendix~\ref{appendixsec:appendiximplements} for more details on implementation. Our codes will be made publicly available after the review process.

\subsection{Profitability and Fairness (RQ1)} \label{subsec:researchquestion1}

\begin{table*}[htbp]
  \small
  \caption{Performance comparison between FairPricing methods and UniformPricing in terms of profitability and fairness under linear and exponential demand models. Standard errors are shown in parentheses. $\uparrow$ indicates that a higher metric value is desired, and $\downarrow$ indicates the opposite.}
  \label{tab:profitfairnessRQ1}
  \resizebox{\textwidth}{!}{
  \begin{tabular}{l|l|cccc|cccc}
    \toprule
    \multicolumn{1}{l}{} & \multicolumn{1}{l}{} & \multicolumn{4}{c}{Pokec-z} & \multicolumn{4}{c}{Pokec-n} \\
    \cmidrule(lr){3-6} \cmidrule(lr){7-10}
    \multicolumn{1}{l}{Demand} & \multicolumn{1}{l}{Method} & $\pi_{\textrm{AVG}}$ ($\uparrow$) & $p_{\textrm{diff}}$ ($\downarrow$) & $\Delta_{\textrm{AVG}}$ ($\downarrow$) & $\eta_{\textrm{AVG}}$ ($\downarrow$) & $\pi_{\textrm{AVG}}$ ($\uparrow$) & $p_{\textrm{diff}}$ ($\downarrow$) & $\Delta_{\textrm{AVG}}$ ($\downarrow$) & $\eta_{\textrm{AVG}}$ ($\downarrow$) \\
    \midrule
    \multirow{5}{*}{Linear} & UniformPricing & 5.90 (0.003) & 0 (0) & 0 (0) & 0 (0) & 5.89 (0.002) & 0 (0) & 0 (0) & 0 (0) \\
    & FairPricing-MLP & 10.21 (0.06) & 0.27 (0.08) & 9.06 (1.09) & -0.09 (0.02) & 9.77 (0.01) & 0.32 (0.09) & 19.47 (0.18) & -0.13 (0.004) \\
    & FairPricing-GCN & 10.85 (0.01) & 0.16 (0.05) & -45.64 (1.03) & -0.76 (0.002) & 10.92 (0.01) & 0.14 (0.04) & -31.84 (0.34) & -0.80 (0.001) \\
    & FairPricing-GAT & 11.38 (0.04) & 0.37 (0.13) & -15.14 (0.99) & -0.57 (0.01) & 11.28 (0.12) & 0.32 (0.08) & -5.48 (0.88) & -0.61 (0.03) \\
    & FairPricing-GraphSAGE & 11.94 (0.03) & 0.43 (0.16) & -8.52 (0.33) & -0.52 (0.01) & 11.65 (0.004) & 0.18 (0.13) & -2.59 (0.29) & -0.58 (0.002) \\
    \midrule
    \multirow{5}{*}{Exponential} & UniformPricing & 7.47 (0.001) & 0 (0) & 0 (0) & 0 (0) & 7.35 (0.002) & 0 (0) & 0 (0) & 0 (0) \\
    & FairPricing-MLP & 10.36 (0.07) & 0.14 (0.06) & 5.19 (0.25) & -0.17 (0.01) & 10.83 (0.01) & 0.18 (0.06) & 12.49 (0.39) & -0.18 (0.003) \\
    & FairPricing-GCN & 13.30 (0.01) & 0.29 (0.09) & -38.36 (0.11) & -0.79 (0.001) & 13.44 (0.02) & 0.20 (0.07) & -28.33 (0.45) & -0.84 (0.003) \\
    & FairPricing-GAT & 13.33 (0.08) & 0.23 (0.10) & -20.63 (1.08) & -0.77 (0.01) & 12.78 (0.41) & 0.21 (0.08) & -13.49 (1.81) & -0.68 (0.06) \\
    & FairPricing-GraphSAGE & 13.25 (0.01) & 0.26 (0.05) & -11.21 (0.27) & -0.69 (0.002) & 13.54 (0.01) & 0.26 (0.03) & -5.17 (0.30) & -0.73 (0.003) \\
    \bottomrule
  \end{tabular}}
\end{table*}

To answer RQ1, we train pricing policies for each dataset using the FairPricing and UniformPricing methods under both demand models, and summarize their evaluation results in Table~\ref{tab:profitfairnessRQ1}. As shown in Table~\ref{tab:profitfairnessRQ1}, all methods based on FairPricing achieve higher profits than UniformPricing, confirming the advantage of personalized pricing in improving profitability. 
As a trade-off, these methods exhibit a certain degree of group-level unfairness compared to the UniformPricing, which attains $p_{\textrm{diff}} = 0$. Nevertheless, under the FairPricing framework, they maintain $p_{\textrm{diff}}$ below the predefined threshold $\tau = 0.5$, and the magnitude of $p_{\textrm{diff}}$ remains considerably smaller than that of the average profit. This result indicates that FairPricing effectively controls price discrimination against protected groups and yields pricing policies that comply with group fairness requirements.

In terms of individual fairness perceptions, all three FairPricing methods with GNN backbones result in negative values of $\Delta_{\textrm{AVG}}$ and $\eta_{\textrm{AVG}}$. This suggests that, under their personalized pricing policies, customers on average experience negative price differences relative to their neighbors and feel positively about receiving favorable offers. These policies improve customer satisfaction compared with UniformPricing, under which all customers remain neutral in perceived unfairness with $\Delta_{\textrm{AVG}} = 0$ and $\eta_{\textrm{AVG}} = 0$.
Interestingly, by replacing the GNN backbone with an MLP, FairPricing-MLP shows positive $\Delta_{\textrm{AVG}}$ and much higher $\eta_{\textrm{AVG}}$ values than its GNN counterparts, indicating worse customer fairness perception. This is reasonable because the network topology information used by GNNs could help capture local relational patterns that promote fair pricing. Consequently, the better fairness perception achieved by the GNN-based methods translates into higher profits than FairPricing-MLP, as positive customer emotions increase demand.

These results confirm the effectiveness of FairPricing in increasing profits and improving individual fairness while maintaining group fairness. Its performance remains consistent across different GNN backbones, with GCN achieving the best individual fairness.

\subsection{Generalization Ability Analysis (RQ2)} \label{subsec:generalizationanalysis}
For FairPricing, a trained pricing policy can be directly applied to an updated network, such as when new customers and connections are added to the retailer's existing network used for training. In our experiments, the testing set represents these newly added customers and connections. To assess the generalization ability of FairPricing, we use a retraining strategy as a benchmark, which retrains a new pricing policy on the combined training and testing sets. A small performance gap between the generalized and retrained policies would indicate that FairPricing generalizes well to network changes.

\begin{figure}[htbp]
  \centering
  \begin{subfigure}[t]{0.49\columnwidth}
    \centering
    \includegraphics[width=\linewidth]{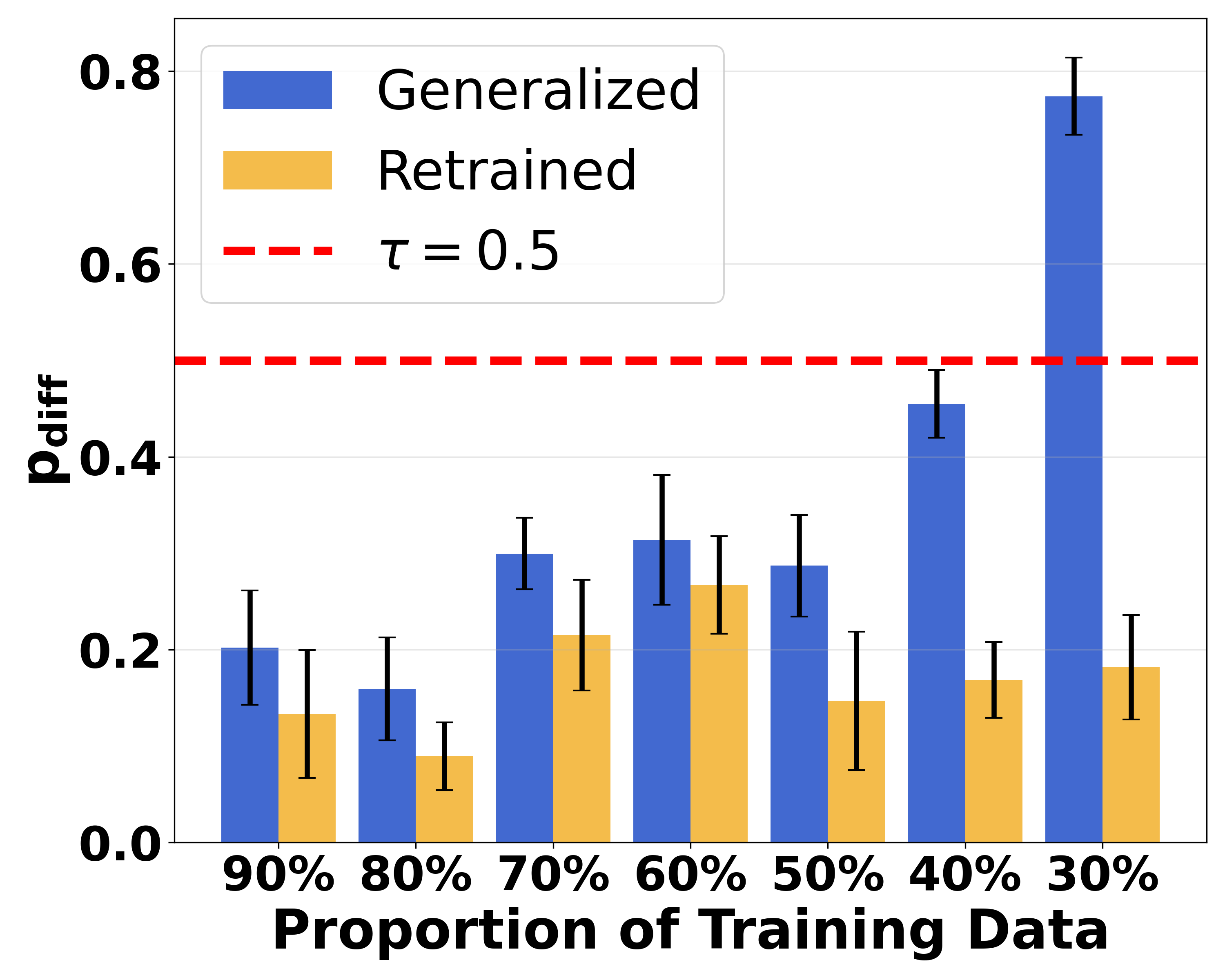}
    \caption{$p_{\textrm{diff}}$}
    \label{fig:generalization_pdiff}
  \end{subfigure}
  \hfill 
  \begin{subfigure}[t]{0.49\columnwidth}
    \centering
    \includegraphics[width=\linewidth]{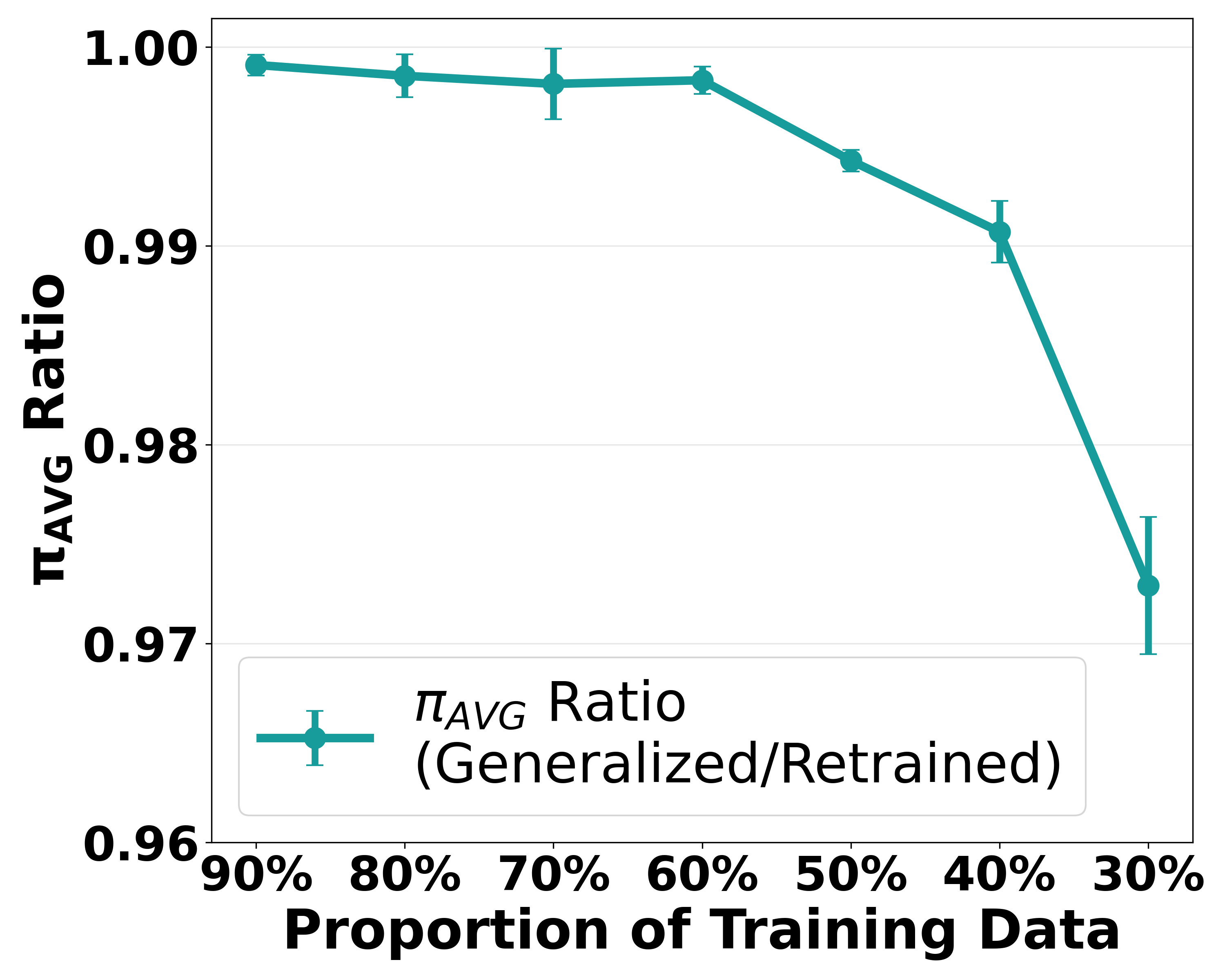}
    \caption{$\pi_{\textrm{AVG}}$ Ratio}
    \label{fig:generalization_piavg}
  \end{subfigure}
\caption{Generalization performance of FairPricing-GCN across varying proportions of training data in terms of $p_{\textrm{diff}}$ and $\pi_{\textrm{AVG}}$.}
\Description{There is a bar plot on the left and a line plot on the right.}
\label{fig:generalization}
\end{figure}

The degree of network changes relative to the training network is a key factor affecting policy generalization ability. To capture this effect, we vary the training data proportion in the random split among $\{90\%, 80\%,\dots,30\%\}$, with the remaining data evenly divided between validation and testing sets. The implementation of the generalized policy follows Section~\ref{subsubsec:implementation}, while the retrained policy is selected as the model achieving the highest profit under $p_{\textrm{diff}} \leq 0.5$. As an illustrative case, we conduct the analysis on Pokec-z under the linear demand, using GCN as the GNN model in FairPricing. The comparisons of $p_{\textrm{diff}}$ and the $\pi_{\textrm{AVG}}$ ratio between the two types of policies are presented in Figures~\ref{fig:generalization_pdiff} and~\ref{fig:generalization_piavg}, respectively, and the results for $\Delta_{\textrm{AVG}}$ and $\eta_{\textrm{AVG}}$ are provided in Appendix~\ref{appendixsec:appendixgeneralization}.

From Figure~\ref{fig:generalization}, we observe that when the training proportion is at least 60\%, the generalized policy achieves comparable group-level fairness on the updated network to the retrained policy, with no statistically significant difference in their $p_{\textrm{diff}}$. As the training proportion decreases below 60\%, the gap in $p_{\textrm{diff}}$ between the two policies gradually widens, and at 30\%, the generalized policy fails to satisfy the $\tau = 0.5$ threshold.
For the ratio of $\pi_{\textrm{AVG}}$ of the generalized policy to that of the retrained policy, the value decreases slightly as the training proportion drops, but remains above 0.97 even at 30\%. At training proportions of 60\% or higher, the ratio exceeds 0.995, indicating that the generalized policy achieves over 99.5\% of the profitability of the retrained policy.
This analysis shows that the pricing policies learned by FairPricing generalize well under moderate network changes, allowing the policy to be directly applied to an updated network. Such generalization enables FairPricing to avoid repeatedly solving optimization problems required by traditional optimization-based methods whenever the network changes.

\subsection{Ablation Study and Hyperparameters (RQ3)}
In FairPricing, both the adversarial debiasing module and the price regularization term are designed to promote group fairness. To analyze their impacts on the performance of FairPricing, particularly regarding $\pi_{\textrm{AVG}}$ and $p_{\textrm{diff}}$, we conduct a set of ablation experiments.
For this analysis, we consider FairPricing-GCN as the base model and construct three incomplete variants:
1) removing the adversarial debiasing module (w/o Adv), 2) removing the price regularization term defined in \eqref{eqn:priceregularization} (w/o Reg), and 3) removing both components (w/o Adv+Reg). We compare their performance on the Pokec-z and Pokec-n datasets under the linear demand, and the results are summarized in Table~\ref{tab:ablationstudyRQ3}. The implementation details for these variants are provided in Appendix~\ref{appendixsec:appendixablation}.

\begin{table}[htbp]
  \small
  \caption{Ablation study of adversarial debiasing and price regularization in FairPricing (results from FairPricing-GCN).}
  \label{tab:ablationstudyRQ3}
  \resizebox{\columnwidth}{!}{
  \begin{tabular}{l|l|cccc}
    \toprule
    \multicolumn{1}{l}{Dataset} & \multicolumn{1}{l}{Method} & $\pi_{\textrm{AVG}}$ ($\uparrow$) & $p_{\textrm{diff}}$ ($\downarrow$) & $\Delta_{\textrm{AVG}}$ ($\downarrow$) & $\eta_{\textrm{AVG}}$ ($\downarrow$) \\
    \midrule
    \multirow{4}{*}{Pokec-z} & FairPricing & 10.85 (0.01) & 0.16 (0.05) & -45.64 (1.03) & -0.76 (0.002) \\
    & w/o Adv & 10.85 (0.01) & 0.34 (0.04) & -45.88 (0.71) & -0.76 (0.002) \\
    & w/o Reg & 10.81 (0.01) & 3.41 (0.20) & -44.19 (0.65) & -0.78 (0.004) \\
    & w/o Adv+Reg & 11.08 (0.01) & 5.80 (0.26) & -50.27 (0.75) & -0.78 (0.003) \\
    \midrule
    \multirow{4}{*}{Pokec-n} & FairPricing & 10.92 (0.01) & 0.14 (0.04) & -31.84 (0.34) & -0.80 (0.001) \\
    & w/o Adv & 10.93 (0.01) & 0.25 (0.07) & -32.47 (0.37) & -0.80 (0.001) \\
    & w/o Reg & 10.89 (0.01) & 1.03 (0.10) & -30.62 (0.55) & -0.79 (0.002) \\
    & w/o Adv+Reg & 11.04 (0.02) & 1.65 (0.12) & -34.71 (0.31) & -0.80 (0.001) \\
    \bottomrule
  \end{tabular}}
\end{table}

According to Table~\ref{tab:ablationstudyRQ3}, w/o Adv+Reg shows a much higher $p_{\textrm{diff}}$ than both w/o Adv and w/o Reg, indicating that the adversarial debiasing module and the price regularization term each contribute meaningfully to mitigating group-level discrimination. Between the two single-component variants, w/o Adv achieves a lower $p_{\textrm{diff}}$ than w/o Reg and better satisfies the $\tau = 0.5$ threshold. This is expected, as the regularization term is defined to directly penalize large disparities in $p_{\textrm{diff}}$.
By comparing w/o Adv with FairPricing-GCN, we observe similar profitability $\pi_{\textrm{AVG}}$, but the $p_{\textrm{diff}}$ of FairPricing-GCN is significantly lower. This suggests that combining both adversarial debiasing and price regularization yields better group fairness without compromising profit. These findings collectively confirm the soundness and effectiveness of the FairPricing design.

The strengths of the price regularization and the adversarial debiasing are determined by the hyperparameters $\lambda$ and $\phi$ in \eqref{eqn:finalobjective}, respectively. To examine the sensitivity of FairPricing to these hyperparameters, we train FairPricing-GCN on Pokec-z under the linear demand as an illustrative example. The optimal values identified by grid search are $\lambda=0.1$ and $\phi=0.01$. We then evaluate performance with respect to $\pi_{\textrm{AVG}}$ and $p_{\textrm{diff}}$ by varying one hyperparameter while keeping the other fixed. The results are presented in Figure~\ref{fig:hyperparameter}, and analysis details can be found in Appendix~\ref{appendixsec:appendixablation}.

\begin{figure}[htbp]
  \centering
  \begin{subfigure}[t]{0.49\columnwidth}
    \centering
    \includegraphics[width=\linewidth]{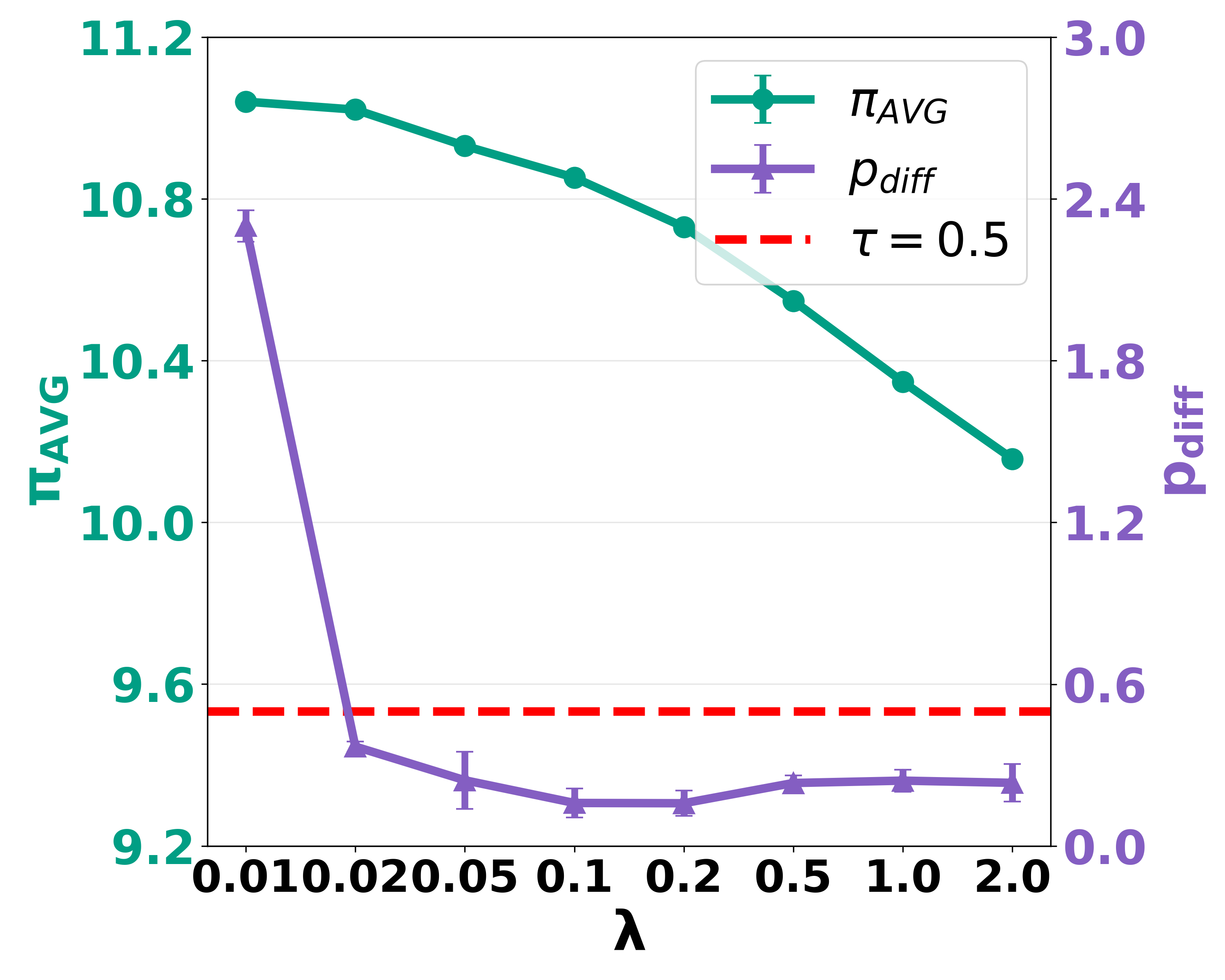}
    \caption{$\lambda$ for $\mathcal{L}_{\textrm{reg}}$}
    \label{fig:hyperparameter_lambda}
  \end{subfigure}
  \hfill 
  \begin{subfigure}[t]{0.49\columnwidth}
    \centering
    \includegraphics[width=\linewidth]{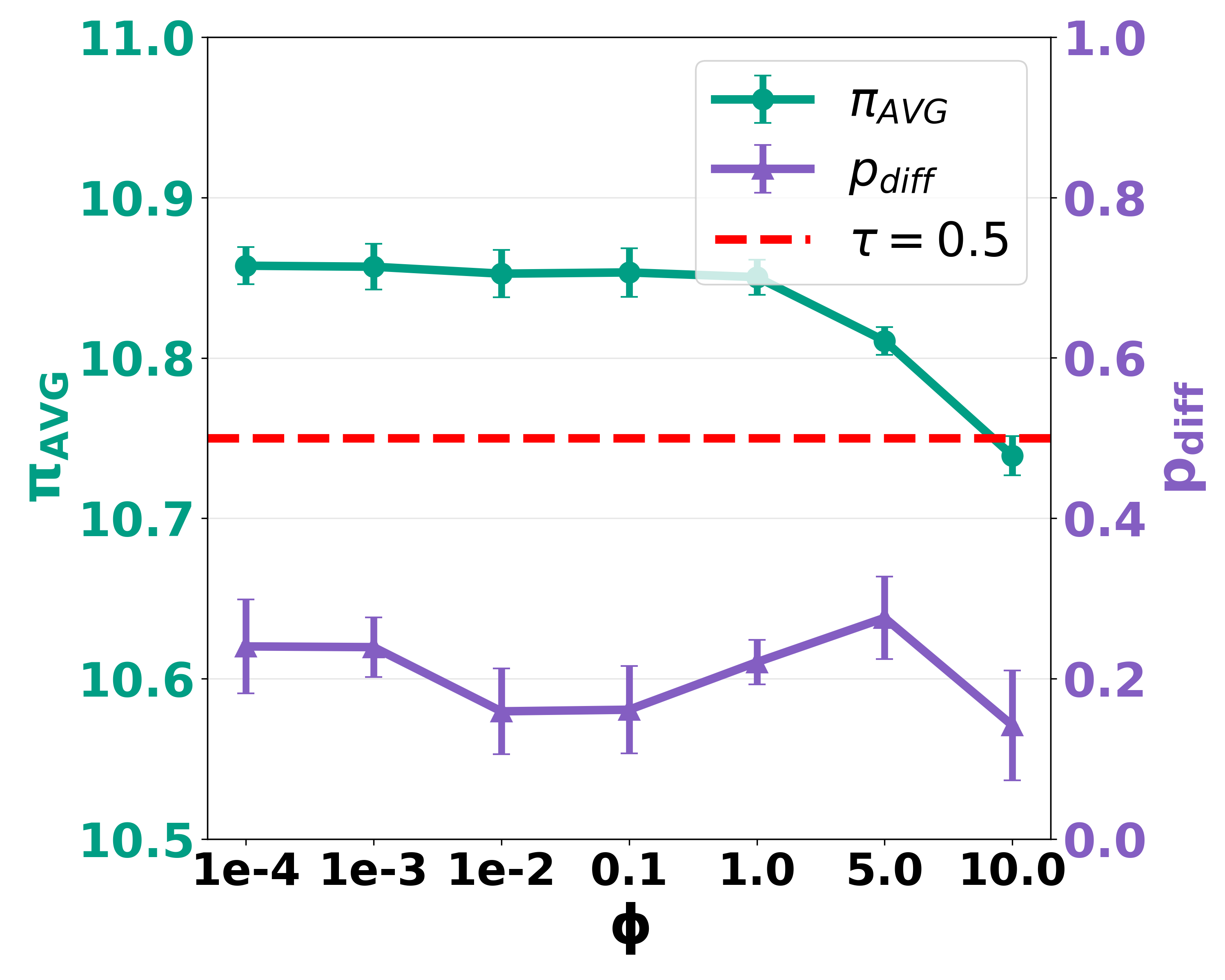}
    \caption{$\phi$ for $\mathcal{L}_{\textrm{adv}}$}
    \label{fig:hyperparameter_phi}
  \end{subfigure}
\caption{Hyperparameter sensitivity analysis.}
\Description{There is a line plot on the left and a line plot on the right.}
\label{fig:hyperparameter}
\end{figure}

Figure~\ref{fig:hyperparameter_lambda} shows that the profitability $\pi_{\textrm{AVG}}$ steadily decreases as $\lambda$ increases, since a larger $\lambda$ imposes stronger regularization on the output prices. The discrimination level $p_{\textrm{diff}}$ is substantially above the $0.5$ threshold when $\lambda = 0.01$. As $\lambda$ increases, $p_{\textrm{diff}}$ drops rapidly at first and then more gradually, and reaches its minimum around $\lambda \in [0.1,0.2]$. For even larger $\lambda$, $p_{\textrm{diff}}$ begins to rise slightly.
As for the impact of $\phi$ shown in Figure~\ref{fig:hyperparameter_phi}, we observe that increasing $\phi$ causes $\pi_{\textrm{AVG}}$ to decrease only slightly at first, but the decline becomes more noticeable once $\phi$ exceeds 1. We also find that as $\phi$ increases, $p_{\textrm{diff}}$ decreases initially and reaches its lowest level around $\phi \in [0.01, 0.1]$. For higher values of $\phi$, $p_{\textrm{diff}}$ exhibits some fluctuations and becomes even smaller when $\phi = 10$. These results suggest that setting $\lambda=0.1$ and $\phi=0.01$ allows FairPricing-GCN to achieve high profitability while improving group fairness.

\subsection{Post-hoc Analysis (RQ4)}
\begin{figure}[htbp]
  \centering
  \includegraphics[width=0.7\columnwidth]{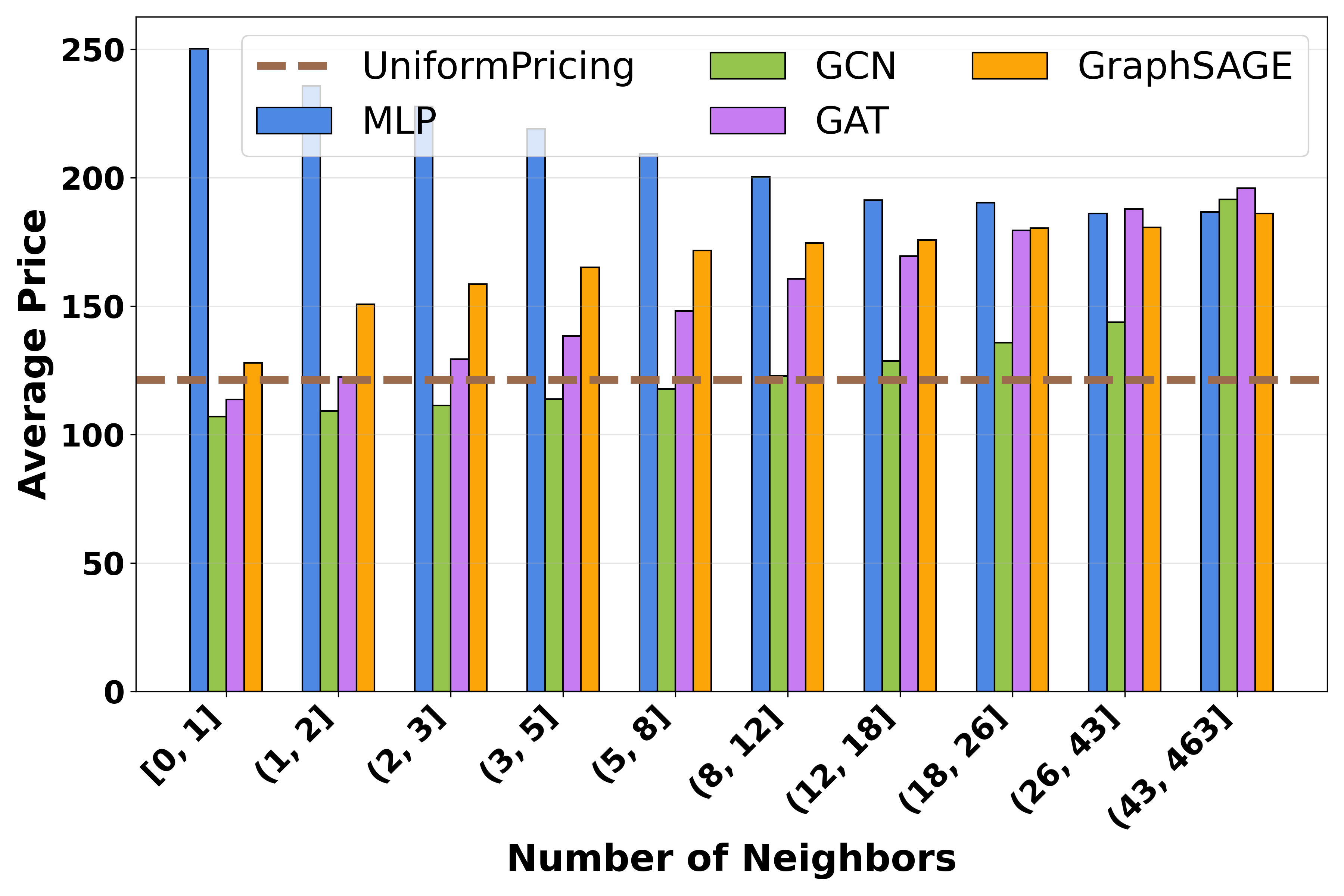}
  \caption{Average prices across customer segments by deciles of the number of neighbors $r_i$. The prefix ``FairPricing-'' is omitted for the four FairPricing variants.}
  \Description{There is a bar plot here.}
  \label{fig:posthocanalysis}
\end{figure}

In Section~\ref{subsec:researchquestion1}, we showed that FairPricing with GNN backbones significantly improves individual fairness perceptions compared to UniformPricing and FairPricing-MLP. To understand how the GNN-based pricing policy achieves this by utilizing network topology, we conduct a post-hoc analysis examining the relationship between its assigned prices and customer network connectivity.
Specifically, we train a policy using each method in Table~\ref{tab:profitfairnessRQ1} on Pokec-z under the linear demand. For every pricing policy, we then divide customers into ten segments based on deciles of the number of neighbors $r_i$ and compute the average price assigned within each segment. The results are shown in Figure~\ref{fig:posthocanalysis}.
We observe that all three GNN-based FairPricing variants exhibit similar pricing patterns: their average assigned prices consistently increase from the lowest decile of $r_i$ to the highest. This contrasts with UniformPricing, which has the same average price across all segments, and with FairPricing-MLP, which assigns higher average prices in the lower deciles of $r_i$.
By assigning higher prices to customers with more neighbors, their surrounding neighbors tend to experience negative price differences, which in turn leads to more favorable fairness perceptions across a larger portion of the population. This strategy enables GNN-based policies to improve individual fairness perceptions on average and offers managerial insights for designing personalized pricing policies that enhance overall customer satisfaction in networked marketplaces.

\section{Conclusion}
In this work, we formulate the personalized pricing problem in social networks with both individual and group fairness considerations. To solve this problem, we propose a novel GNN-based framework named FairPricing that learns feature-based personalized pricing policies using customer features and network topology.
Experimental results show that FairPricing enables retailers to achieve high profitability while improving individual fairness perceptions and satisfying group fairness requirements. The results also confirm that the learned pricing policy generalizes well under moderate network changes, enhancing its practical applicability.
Future research could extend the problem formulation to account for competition among multiple retailers in the market or explore alternative group fairness measures that capture broader societal objectives, such as consumer surplus and social welfare.



\bibliographystyle{ACM-Reference-Format}
\bibliography{reference}

\appendix

\section{Proof of Proposition~\ref{thm:advlearning}} \label{appendixsec:proofofprop}
In this section, we provide the proof of Proposition~\ref{thm:advlearning}.

\begin{proof}
    Following the conclusions of Proposition 1 in \cite{goodfellow2014generative}, for a fixed $\psi_G$, the optimal adversary $\psi_A$ that maximizes $V(\psi_G,\psi_A)$ is
    \begin{equation*}
        \psi_{A|\psi_G}^*(\mathbf{h}^{(K)}) = \frac{f(\mathbf{h}^{(K)} | s = 1)}{f(\mathbf{h}^{(K)} | s = 1) + f(\mathbf{h}^{(K)} | s = 0)}.
    \end{equation*}
    By substituting $\psi_{A|\psi_G}^*$ into $V(\psi_G,\psi_A)$, the minimax game in \eqref{eqn:advvaluationfunction} can be reformulated as minimizing $V(\psi_G,\psi_{A|\psi_G}^*)$, which equals
    \begin{align*}
        & \mathrm{E}_{\mathbf{h}^{(K)} \sim f(\mathbf{h}^{(K)} | s = 1)} [ \log \frac{f(\mathbf{h}^{(K)} | s = 1)}{f(\mathbf{h}^{(K)} | s = 1) + f(\mathbf{h}^{(K)} | s = 0)} ] \\
        +\; & \mathrm{E}_{\mathbf{h}^{(K)} \sim f(\mathbf{h}^{(K)} | s = 0)} [\log \frac{f(\mathbf{h}^{(K)} | s = 0)}{f(\mathbf{h}^{(K)} | s = 1) + f(\mathbf{h}^{(K)} | s = 0)}] \\
        = & -\log(4) + 2 JSD(f(\mathbf{h}^{(K)} | s = 1) \| f(\mathbf{h}^{(K)} | s = 0)),
    \end{align*}
    where $JSD(f(\mathbf{h}^{(K)} | s = 1) \| f(\mathbf{h}^{(K)} | s = 0))$ is the Jensen–Shannon divergence between the two conditional distributions and attains its minimum value of 0 if and only if the two distributions are identical. Thus, the global minimum of $V(\psi_G,\psi_A)$ is achieved when $f(\mathbf{h}^{(K)} | s = 1) = f(\mathbf{h}^{(K)} | s = 0)$, $\forall \, \mathbf{h}^{(K)}$. If this equality holds, it follows directly that $f_{\psi_P}(p | s = 1) = f_{\psi_P}( p | s = 0)$ for any pricing layer function $\psi_P$. This completes the proof.
\end{proof}

\section{Experimental Details}
\subsection{Datasets} \label{appendixsec:appendixdatasets}
For each dataset, we treat gender as the protected attribute $s$ and randomly select ten features from the remaining user attributes as the non-protected attributes $\mathbf{x}$. The basic statistics and selected attributes of Pokec-z and Pokec-n are summarized in Table~\ref{tab:datasetsummary}.
In each dataset, considering a monopolist selling a single product to the users, we assume that the ten selected non-protected attributes $\mathbf{x}_i$ together with the gender $s_i$ determine a customer's willingness to pay $u_i$ for the product.

\begin{table*}
  \caption{The statistics and attributes of datasets (attribute names translated from Slovak to English).}
  \label{tab:datasetsummary}
  \begin{tabular}{ccc}
    \toprule
    Dataset & Pokec-z & Pokec-n \\
    \midrule
    \# of nodes & 67,796 & 66,569 \\
    \# of edges & 882,765 & 729,129 \\
    Protected attribute $s$ & \textit{gender} & \textit{gender} \\
    \multirow{5}{*}{Non-protected attributes $\mathbf{x}$} & \textit{region}, \textit{AGE}, \textit{hobbies\_friends} & \textit{completion\_percentage}, \textit{education\_high\_school} \\
    & \textit{hobbies\_sports}, \textit{hobbies\_party} & \textit{marital\_single}, \textit{I\_want\_to\_have\_children} \\ 
    & \textit{hobbies\_dancing}, \textit{hobbies\_visiting\_museums} & \textit{hobbies\_web\_surfing}, \textit{hobbies\_PC\_games} \\
    & \textit{hobbies\_housework}, \textit{education\_university} & \textit{smoke\_regularly}, \textit{drink\_occasionally} \\
    & \textit{like\_movies\_comedy} & \textit{region}, \textit{hobbies\_reading} \\
  \bottomrule
\end{tabular}
\end{table*}

\subsection{Simulation Settings for Demand Models} \label{appendixsec:appendixsimulation}
The willingness to pay $u_i$ of each customer follows a distribution with cumulative distribution function $F_{\tilde{\mathbf{x}}_i}(\cdot)$ that is known by the retailer. For the linear and exponential demand models considered, $F_{\tilde{\mathbf{x}}_i}(\cdot)$ is characterized by a distribution parameter $g(\tilde{\mathbf{x}}_i)$, and thus the function $g(\cdot)$ is assumed to be known. In our experiments, we conduct simulation studies by specifying the functional form of $g(\tilde{\mathbf{x}}_i)$. Using the attributes for each dataset given in Table~\ref{tab:datasetsummary}, the detailed simulation settings of $g(\tilde{\mathbf{x}}_i)$ under each demand model are provided below.

\paragraph{Pokec-z} Under the linear demand model, we first define a linear function of attributes $t(\tilde{\mathbf{x}}_i)$, given by
\begingroup
\small
\begin{equation*}
\begin{aligned}
    &t(\tilde{\mathbf{x}}_i) = 80 + 25region - AGE + 25hobbies\_friends + \\
    &\; 25hobbies\_sports + 25hobbies\_party + 25hobbies\_dancing - \\
    &\; 10hobbies\_visiting\_museums - 10hobbies\_housework + \\
    &\; 25education\_university + 25like\_movies\_comedy + 60gender,
\end{aligned}
\end{equation*}
\endgroup
and then we set $g(\tilde{\mathbf{x}}_i) = \max(0, t(\tilde{\mathbf{x}}_i))$.
For the exponential demand model, we begin by defining $t(\tilde{\mathbf{x}}_i)$ as
\begingroup
\small
\begin{equation*}
\begin{aligned}
    &t(\tilde{\mathbf{x}}_i) = -0.8 - 0.2region +0.042 AGE - 0.2hobbies\_friends - \\
    &\; 0.2hobbies\_sports - 0.2hobbies\_party - 0.2hobbies\_dancing + \\
    &\; 0.1hobbies\_visiting\_museums + 0.1hobbies\_housework - \\
    &\; 0.2education\_university - 0.2like\_movies\_comedy - 0.6gender.
\end{aligned}
\end{equation*}
\endgroup
To ensure that the range of the expected willingness to pay $\mathrm{E}(u_i)$ across all customers is comparable to that under the linear demand model, we let $g(\tilde{\mathbf{x}}_i) = 0.008 + 0.022 \sigma(t(\tilde{\mathbf{x}}_i))$, where $\sigma(\cdot)$ denotes the sigmoid function.

\paragraph{Pokec-n} In the Pokec-n dataset, we follow a similar procedure as above to specify $g(\tilde{\mathbf{x}}_i)$. Under the linear demand model, $t(\tilde{\mathbf{x}}_i)$ is defined as
\begingroup
\small
\begin{equation*}
\begin{aligned}
    &t(\tilde{\mathbf{x}}_i) = 60 + completion\_percentage + 15education\_high\_school + \\
    &\; 15marital\_single + 20I\_want\_to\_have\_children + \\
    &\; 20hobbies\_web\_surfing + 20hobbies\_PC\_games - \\
    &\; 20smoke\_regularly - 20drink\_occasionally - 20region - \\
    &\; 20hobbies\_reading + 50gender,
\end{aligned}
\end{equation*}
\endgroup
and $g(\tilde{\mathbf{x}}_i) = \max(0, t(\tilde{\mathbf{x}}_i))$. As for the exponential demand model, we set
\begingroup
\small
\begin{equation*}
\begin{aligned}
    &t(\tilde{\mathbf{x}}_i) = -0.6 - 0.02completion\_percentage - \\ 
    &\; 0.15education\_high\_school - 0.15marital\_single - \\
    &\; 0.2I\_want\_to\_have\_children - 0.2hobbies\_web\_surfing - \\
    &\; 0.2hobbies\_PC\_games + 0.2smoke\_regularly + \\
    &\; 0.2drink\_occasionally + 0.2region + 0.2hobbies\_reading - 0.5gender,
\end{aligned}
\end{equation*}
\endgroup
and $g(\tilde{\mathbf{x}}_i) = 0.008 + 0.052 \sigma(t(\tilde{\mathbf{x}}_i))$.

In addition, based on these specifications of $g(\tilde{\mathbf{x}}_i)$, we examine the distribution of $\mathrm{E}(u_i)$ across customers in each dataset to determine reasonable values of the upper pricing bound $p_{\textrm{max}}$ and the cost $c$ for our simulation studies. Accordingly, we set $p_{\textrm{max}}=400$ and $c=80$ for Pokec-z, and $p_{\textrm{max}}=350$ and $c=70$ for Pokec-n.

\subsection{Implementation Details} \label{appendixsec:appendiximplements}
In this section, we first present the details of the GNN architecture implemented for each of the three GNN-based FairPricing variants.

\textbf{FairPricing-GCN}. We adopt a GCN with one hidden layer of dimension 128, and set the output dimension $q=128$ as well.

\textbf{FairPricing-GAT}. We use a GAT with one hidden layer and set hidden\_dim=128 and the output dimension $q=128$. The number of attention heads in the output layer is 1. In the hidden layer, we use 4 heads for Pokec-z under the linear demand, 8 heads for Pokec-z under the exponential demand, and 10 heads for Pokec-n under both demand models. The LeakyReLU negative slope is set to 0.05.

\textbf{FairPricing-GraphSAGE}. We use a one-hidden-layer GraphSAGE with hidden\_dim=128 and $q=128$. The aggregator type is the mean aggregator.

We also apply dropout with a rate of 0.1 to these variants during training. For FairPricing-MLP, we use an MLP that has one hidden layer and hidden\_dim=128. The adversary $\psi_A$ is implemented as a linear classifier. We adopt the Adam optimizer for all experiments, with a learning rate of 0.001 and a weight decay of 0.0005.

By setting the group-level fairness threshold in \eqref{eqn:fairpricingdef} to $\tau = 0.5$, for each implemented FairPricing variant, we select the model with the highest profitability $\pi_{\textrm{AVG}}$ among all candidate models that satisfy $p_{\textrm{diff}} \leq 0.5$. This model selection with early stopping is based on the combined training and validation sets. As for UniformPricing, since it always satisfies this threshold with $p_{\textrm{diff}} = 0$, we determine the optimal uniform price within $[0, p_{\textrm{max}}]$ via a grid search with step size 0.1, selecting the price that yields the highest profitability.

For the FairPricing variants, the hyperparameters $\lambda$ and $\phi$ are selected based on cross-validation with grid search for each demand model in each dataset. Specifically, we vary $\lambda \in \{0.1, 0.2, 0.5, 1, 2\}$ and $\phi \in \{0.001, 0.01, 0.1, 1\}$. For FairPricing-GCN, $\lambda=0.1$ and $\phi=0.01$ are set across both datasets and demand models. For FairPricing-GAT, the chosen values are: $\lambda=0.1$, $\phi=1$, Pokec-z linear demand; $\lambda=0.2$, $\phi=0.01$, Pokec-z exponential demand; $\lambda=0.1$, $\phi=0.1$, Pokec-n linear demand; $\lambda=0.5$, $\phi=1$, Pokec-n exponential demand. For FairPricing-GraphSAGE, we set $\lambda=0.1$ and $\phi=0.1$ across both datasets and demand models. For FairPricing-MLP, the chosen values are: $\lambda=0.1$, $\phi=0.1$, Pokec-z linear demand; $\lambda=0.5$, $\phi=0.1$, Pokec-z exponential demand; $\lambda=0.1$, $\phi=0.1$, Pokec-n linear demand; $\lambda=0.1$, $\phi=1$, Pokec-n exponential demand.

\subsection{Generalization Ability Analysis} \label{appendixsec:appendixgeneralization}
\begin{figure}[htbp]
  \centering
  \begin{subfigure}[t]{0.49\columnwidth}
    \centering
    \includegraphics[width=\linewidth]{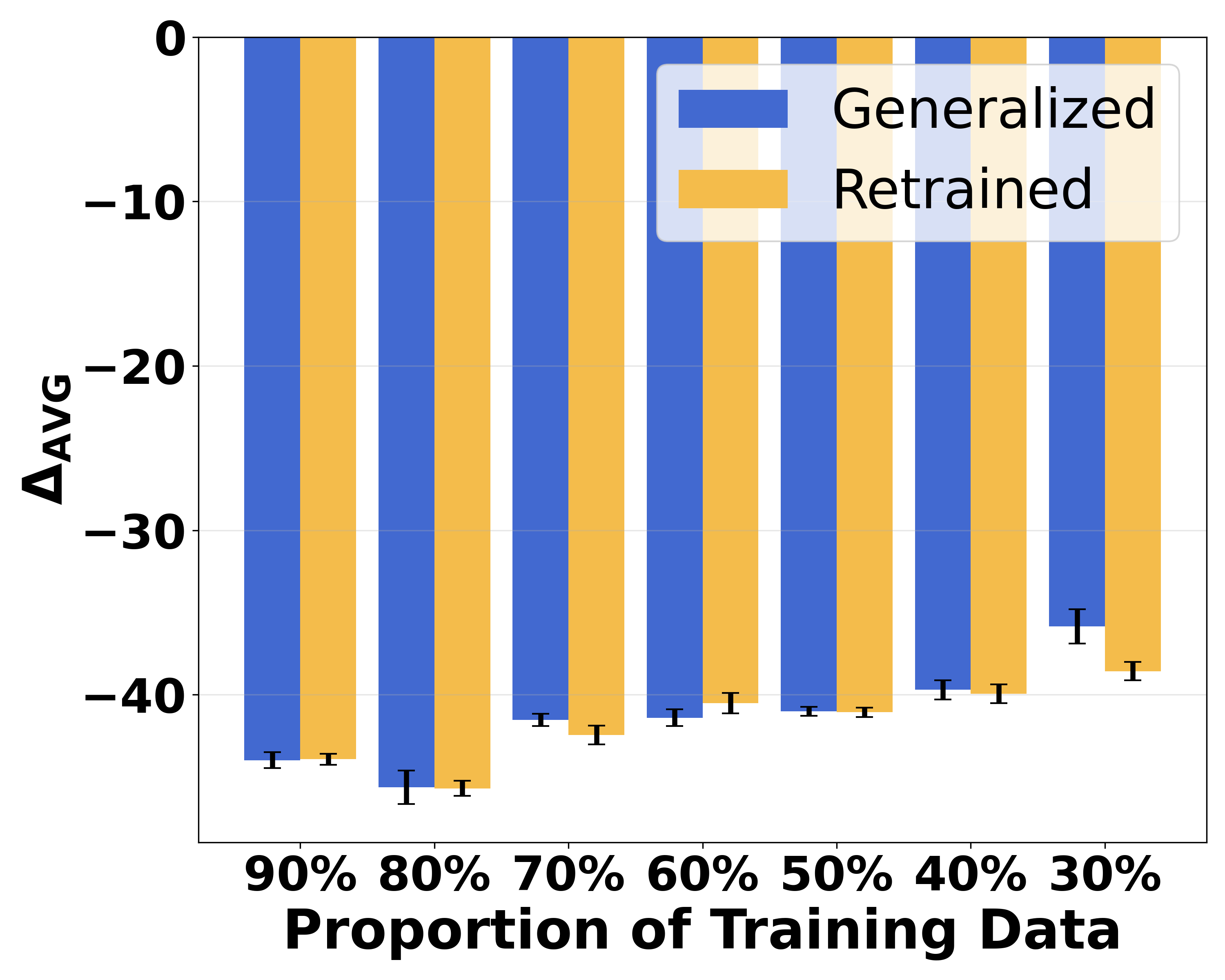}
    \caption{$\Delta_{\textrm{AVG}}$}
    \label{fig:generalization_indpridiff}
  \end{subfigure}
  \hfill 
  \begin{subfigure}[t]{0.49\columnwidth}
    \centering
    \includegraphics[width=\linewidth]{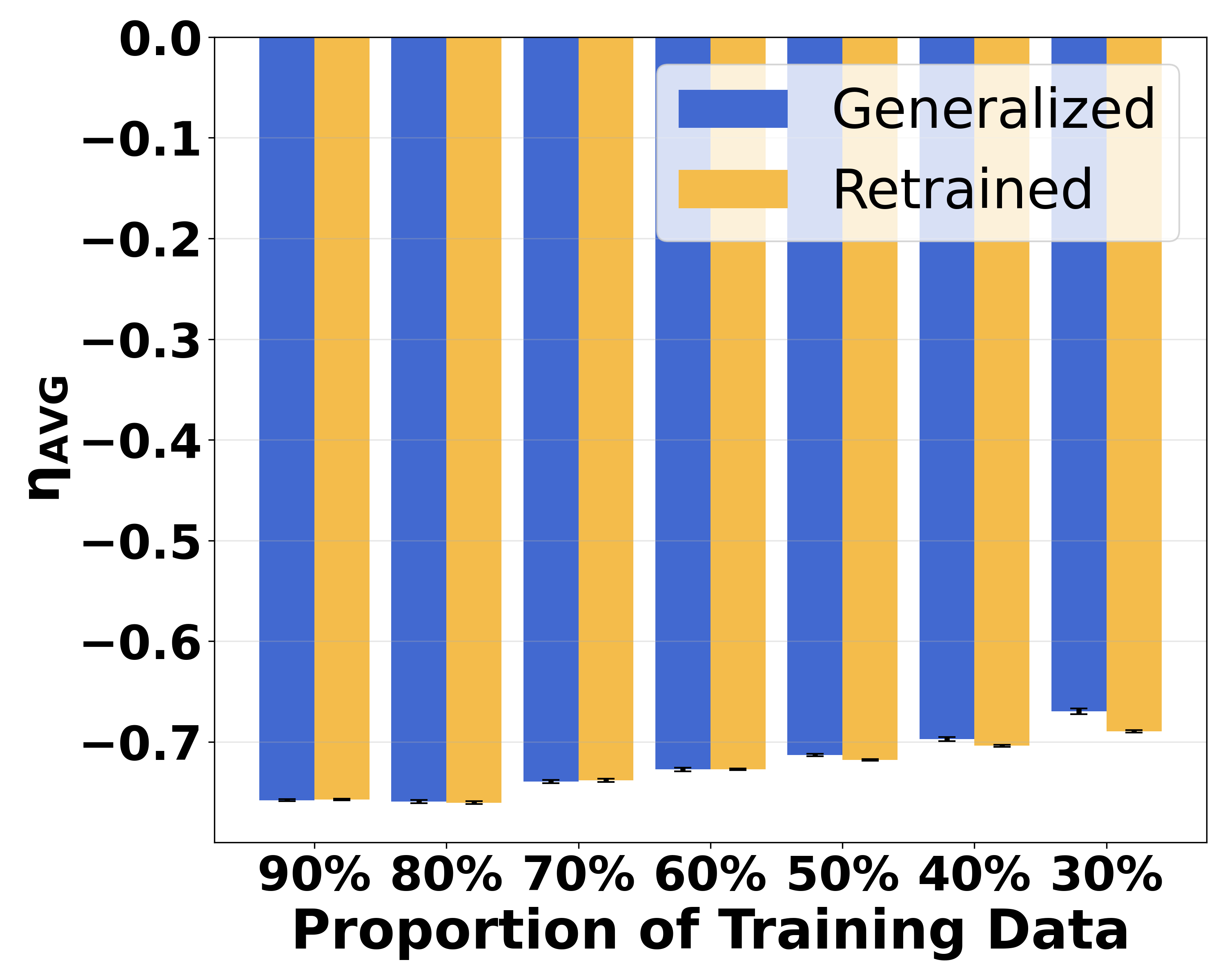}
    \caption{$\eta_{\textrm{AVG}}$}
    \label{fig:generalization_indeta}
  \end{subfigure}
\caption{Generalization performance of FairPricing-GCN across varying proportions of training data in terms of $\Delta_{\textrm{AVG}}$ and $\eta_{\textrm{AVG}}$.}
\Description{There is a bar plot on the left and a bar plot on the right.}
\label{fig:generalization_appendix}
\end{figure}

In the generalization ability analysis, we train the generalized policy on the training set and follow the procedures in Appendix~\ref{appendixsec:appendiximplements} to perform model selection with early stopping based on the combined training and validation sets. The trained policy is then applied to the combined training and testing sets to assign prices using customer features and the new network structure as input, and we evaluate the generalized policy based on these assigned prices. To provide a benchmark, we retrain a new pricing policy on the combined training and testing sets and select the model that achieves the highest profitability while satisfying $p_{\textrm{diff}} \leq 0.5$. This represents a scenario in which the retailer solves a new pricing policy whenever the network changes, similar to optimization-based methods that require re-optimization after each network update. The comparisons of $p_{\textrm{diff}}$ and $\pi_{\textrm{AVG}}$ between the generalized and retrained policies have been shown in Figure~\ref{fig:generalization}, and their differences in terms of the individual fairness perception measures $\Delta_{\textrm{AVG}}$ and $\eta_{\textrm{AVG}}$ are presented in Figure~\ref{fig:generalization_appendix}.

Figure~\ref{fig:generalization_appendix} indicates that when the training proportion is at least 60\%, the generalized policy achieves individual fairness perception levels similar to those of the retrained policy, with no statistically significant differences in either $\Delta_{\textrm{AVG}}$ or $\eta_{\textrm{AVG}}$. As the training proportion becomes less than 60\%, their $\Delta_{\textrm{AVG}}$ values remain statistically comparable, while the generalized policy exhibits a slightly higher $\eta_{\textrm{AVG}}$. Finally, at a proportion of 30\%, the generalized policy results in worse individual fairness outcomes, as reflected by significantly higher $\Delta_{\textrm{AVG}}$ and $\eta_{\textrm{AVG}}$. These findings support the conclusion in Section~\ref{subsec:generalizationanalysis} that the pricing policies learned by FairPricing generalize well under moderate network changes from the perspective of individual fairness measures.

\subsection{Ablation Study and Hyperparameters} \label{appendixsec:appendixablation}
In the ablation experiments, for the incomplete variant w/o Adv, the hyperparameter $\lambda$ is selected via cross-validation with a grid search over $\{0.05,0.1,0.2,0.5,1,2,5,10,20\}$ for each dataset. For the incomplete variant w/o Reg, the hyperparameter $\phi$ is tuned over $\{0.001,0.01,0.1,0.2,0.5,1,2,5,10,20\}$.
Since removing the price regularization term makes it difficult for w/o Reg to meet the predefined threshold $\tau = 0.5$, during model selection with early stopping, we select the model that achieves the lowest group unfairness level $p_{\textrm{diff}}$ while maintaining a profitability $\pi_{\textrm{AVG}}$ similar to that of FairPricing-GCN and w/o Adv.

For the hyperparameter sensitivity analysis, since the optimal values identified by grid search are $\lambda=0.1$ and $\phi=0.01$, we fix $\phi=0.01$ and vary $\lambda$ to evaluate the performance of FairPricing-GCN in Figure~\ref{fig:hyperparameter_lambda}, while in Figure~\ref{fig:hyperparameter_phi} we fix $\lambda=0.1$ and vary $\phi$. The group-level fairness threshold remains $\tau = 0.5$ in this analysis for model selection with early stopping, except when $\lambda = 0.01$ and $\phi=0.01$, where FairPricing-GCN struggles to satisfy the threshold. In this case, we treat $\tau$ as a hyperparameter and tune its value while keeping $\lambda = 0.01$ and $\phi=0.01$ fixed.

\end{document}